\documentclass[12pt,preprint]{aastex}

\shorttitle{Wave-Like Phenomena in the Disrupting Magnetic Field}
\shortauthors{Wang et al.}

\begin{document}

\title{Numerical Experiments of Wave-Like Phenomena Caused by the Disruption of an
Unstable Magnetic Configuration}

\author{Hongjuan Wang\altaffilmark{1, 2},
Chengcai Shen\altaffilmark{1, 2} and Jun Lin \altaffilmark{1, 3}}

\altaffiltext{1}{National Astronomical Observatories of China/Yunnan
Astronomical Observatory, Chinese Academy of
              Sciences, Kunming, Yunnan 650011, China}
\altaffiltext{2}{Graduate School of Chinese Academy of Sciences,
Beijing 100039, China}
\altaffiltext{3}{Harvard-Smithsonian Center for Astrophysics, 60
Garden Street, Cambridge, MA 02138, USA}

\begin{abstract}
The origin of the Moreton wave observed in the chromosphere and the EIT wave observed in the corona during the eruption remains being an active research subject for a while. We investigate numerically in this work the evolutionary features of the magnetic configuration that includes a current-carrying flux rope,
which is used to model the filament, after the loss of equilibrium in the system takes place in a catastrophic fashion. Rapid motions of the flux rope following the catastrophe invokes the velocity vortices behind the rope, and may invoke as well slow and fast mode shocks in front of the rope. The velocity vortices at each side of the flux rope propagate roughly horizontally away from the area where it is produced, and both shocks expand toward the flank of the flux rope. The fast one may eventually reach the bottom boundary and produces two echoes moving back into the corona, but the slow one and the vortices totally decay somewhere in the lower corona before arriving the bottom boundary. The interaction of the fast shock with the boundary leads to disturbance that accounts for the Moreton wave observed in H$\alpha$, and the disturbance in the corona caused by the slow shock and the velocity vortices should account for the EIT wave whose speed is about 40\% that of the Moreton wave. Implication of these results to the observed correlation of the type II radio burst to the fast and the slow mode shocks, and that of EIT waves to CMEs and flares have also been discussed.
\end{abstract}


\section{Introduction}
Solar flare, eruptive prominence, and coronal mass ejection (CME)
constitute the most significant manifestations in the violent energy
conversion process in the solar system, which is also known as the
solar eruption. In a typical eruption, a huge amount of energy (up
to 10$^{32}$~ergs), magnetic flux (up to $10^{22}$~Mx), and plasma
(up to $10^{16}$~g) is flowing into the outermost corona and
interplanetary space (Zhang \& Low 2005). Meanwhile, many activities
due to the secondary effect of the energy conversion are invoked
(Chen et al. 2006, 2007 and references therein). These activities
include various types of radio bursts, the Moreton wave sweeping the
chromosphere, the EIT wave, the X-ray wave, and the coronal dimming
propagating in the corona, and so on (e.g., see Table 1 of Hudson \&
Cliver 2001, and Forbes et al. 2006 for a brief review). It is now
widely accepted that radio bursts, such as types II and III radio
bursts, are caused by the energetic electrons accelerated in the
CME-driven shock and in the flare region, respectively (e.g., see
Bastian et al. 1998), and that the EIT dimming seen by the
Extreme-Ultraviolet (EUV) Imaging Telescope (EIT, Delaboudini\`{e}re
et al. 1995) on board the $Solar$ $and$ $Heliospheric$ $Observatory$
($SOHO$), results from the sudden depletion in the plasma density in
a rapid energy release process. As for the Moreton wave and the EIT
wave, it is very likely that they are the results of CMEs, but their
relationship and the nature of EIT wave are still of the subject of
active research, and no hard and fast conclusion could yet be drawn.
Furthermore, the debate on the relationship among various global
waves (Moreton, EIT, X-ray waves) is still far from being settled.

The Moreton wave was identified for the first time by Moreton \&
Ramsey (1960) in several flare events. They noticed the disturbance
of the chromosphere observed in H$\alpha$ propagating distances over
$10^{5}$ km at speeds ranging from 400 to 2000 km~s$^{-1}$ (e.g.,
see also Becker 1958 and Smith \& Harvey 1971). Therefore, propagation of the
Moreton wave is always super-Alfv\'{e}nic. Observations indicate
that the Moreton wave also has co-spatial signatures in the corona
known as the X-ray disturbance or wave (e.g., see Rust \&
\v{S}vestka 1979; Khan \& Aurass 2002; Narukage et al. 2002), and
the lifetime of the X-ray wave is usually less than 10 minutes.

For the first time, Rust \& \v{S}vestka (1979) noticed the
propagating phenomenon in soft X-ray when studying an event that was
observed by the Skylab. This event caused a coronal disturbance seen
in soft X-ray propagating at speed up to 400 km~s$^{-1}$. They
pointed out that the disturbance resulted from the disruption of a
filament. Because this filament eruption preceded the associated
flare, the source of the disturbance was unlikely to be the flare.
In the era of the Skylab, the term ``CME" did not exist yet. Based
on our present knowledge on the flare, eruptive prominence, and CME,
we believe that the event observed by Rust \& \v{S}vestka (1979)
started with an eruptive prominence and produced a CME consequently,
and the CME ignited the disturbance and the consequent propagating phenomenon seen in soft X-ray.

Khan \& Aurass (2002) reported the appearance of the Moreton wave
and the simultaneous wave-like disturbance in the corona. Narukage
et al. (2004) also noticed that associated with the disturbance in
the chromosphere, the corona showed disturbance in soft X-ray at the
same time. They named such a disturbance X-ray wave. They found that
the disturbance in two layers propagate roughly at same speed, such
that the Moreton wave propagated at speed of 500 km~s$^{-1}$ and the
X-ray wave at speed of 600 km~s$^{-1}$. They suggested that the
X-ray wave is a weak fast mode shock in the corona, and thus the
coronal counterpart of the Moreton wave. Uchida (1968) ascribed the
Moreton wave to the coronal fast-mode shock front as the wave skirt
sweeps the chromosphere, and further pointed out that the fast-mode
shock wave could invoke the type II radio burst as well (Uchida
1974). The tight association of the type II radio bursts to the
Moreton wave (e.g., see Wild et al. 1963; Kai 1970; Uchida 1974;
Pinter 1977; Chen et al. 2005a) strongly suggests the shock origin
of the Moreton wave. Cliver et al. (1999) also presented evidence
based on the SMM and the GOES X-ray data, and the \emph{Solwind}
coronagraph data that type II radio bursts and the Moreton waves
have their root cause in fast CMEs.

The CME-driven shock is usually believed to account for the type II
radio bursts (e.g., see Mancuso et al. 2002; Mancuso \& Raymond
2004; Lin et al. 2006; and references therein). It has been well
established that km type II bursts are manifestations of
interplanetary shocks formed in front of CMEs through its
piston-driven mechanism (Sheeley et al. 1985), and that dm-hm type
II bursts generally correspond to the propagation of CME-driven
shocks through the corona. The origin of metric type II bursts is
still an open question. There are two competing classes of models
for the metric type II radio burst: those that assuming the
CME-driven shock origin and those that flare-related blast wave
origin (e.g., see also Gopalswamy \& Hanaoka 1998; Lin et al. 2006; and references therein).

Cliver et al. (2005) re-examined the relationship between the rapid
acceleration phase of CMEs and the onset of type II bursts for six
events that figured prominently in the debate on the ``flare versus
CME origin of the metric type II radio bursts." Each of these events
had been investigated by one or more sets of authors with the
general conclusion that flare blast waves or flare ejecta were
responsible for the metric type II radio bursts. But Cliver et al.
(2005) arrived at the opposite conclusion that ``CMEs remain a
strong candidate to be principal/sole driver of metric type II
shocks vis-\`{a}-vis flare blast waves/ejecta", and ``these six
events exhibited ample evidence of dynamic behavior" of several
important aspects that ``was consistent with the cataclysmic
disruption of the low solar atmosphere one would expect to be
associated with CMEs." The result of Lin et al. (2006) also provided
a very positive argument in theory for the CME-driven shock origin
of the type II burst.

The more debatable question currently, as well as another important
focus of this work, is the so-called EIT wave, which was, for the
first time, directly imaged by EIT in 195 {\AA}. It was seen to
propagate global disturbances in the corona, which are known as the
EIT wave later on. The EIT wave is most visible in the lower corona
(at 1-2 MK), its lifetime is over an hour, and the average velocity
ranges from 25 to 450 km~s$^{-1}$ (Thompson \& Myers, 2009), which
is about a third or less of those of Moreton waves. Since in some
events a sharp EUV wave front was found to be cospatial with an
H$\alpha$ Moreton wave (Thompson et al. 1998,1998, 2000) and with a
soft X-ray wave in the corona (Khan \& Aurass, 2002; Hudson et al.,
2003), EIT waves were thus considered as the coronal counterparts of
the chromospheric Moreton waves.

Here we have a problem: some authors considered the X-ray wave the
coronal counterpart of the Moreton Wave, and the others considered
the EIT wave the counterpart. These two views actually constitute
part of the debate in the community on the relationship among these
different global wave phenomena.

By investigating three different EIT waves, Delann\'{e}e (2000)
suggested that the EIT wave is more closely related to the magnetic
field evolution involved in CMEs than to wave propagation driven by
solar flares. Ballai et al. (2005) studied $TRACE$ EUV data to show
that EIT waves are indeed waves with a well-defined period. Based on
the MHD simulations, Chen et al. (2002, 2005a, 2005b) showed that
EIT waves are thought to be formed by successive stretching or
opening of closed field lines driven by an erupting flux rope as
originally suggested by Delann\'{e}e \& Aulanier (1999). However,
the arguments of Wills-Davey et al. (2007) constitute the main
difficulties of Chen et al. (2002, 2005a, 2005b) in ascribing the
EIT wave to the opening of the closed field lines, and suggest that
the origin and the nature of EIT waves remain elusive, and more
investigations need to be done.

Recently, Tripathi \& Raouafi (2007) studied a CME event on 2000
March 5, and provided strong evidence in favor of the interpretation
that EIT waves are indeed a counterpart of the CME-driven shock wave
in the lower corona. Attrill et al. (2007) investigated properties
of two EIT waves and noticed a new property of EIT waves: dual
brightenings. They suggested that a mechanism where driven magnetic
reconnections between the skirt of the expanding CME magnetic field
and quiet-Sun magnetic loops generate the observed bright diffuse
front, and the dual brightenings and the widespread diffuse dimming
are identified as innate characteristics of this process.

In the spirit of previous works, we are studying in the present work
the origin of various wave-like phenomena mentioned above via MHD
numerical experiments. We shall use a well known numerical code,
ZEUS-2D, that was originally developed by Stone \& Norman (1992a,
1992b) and Stone et al. (1992) and is open to the public, to perform
our investigations. We describe the numerical approaches, including
the MHD equations, and the boundary and initial conditions applied to
the simulation in the next section. Our results are presented in Section
3, we present our discussions on these results in Section 4, and
finally we summarize this work in Section 5.

\section{Formulae and Numerical Approaches}
We consider a two-dimensional magnetic configuration in the
semi-infinite $x$-$y$ plane with $y=0$ being the bottom boundary
located on the bottom of the photosphere or the top of the
chromosphere, and $y>0$ corresponding to the chromosphere and the
corona. This model treats the current-carrying filament floating in
the corona as a force-free flux rope that is located at height $h$
on the $y$-axis. The background field is represented by a line
dipole in the photosphere at depth $d$ below the boundary $y=0$.
This work is a follow-up to that of Forbes (1990). Evolution in this
magnetic system is governed by the following two-dimensional,
time-dependent, ideal MHD equations in the Cartesian coordinates:
\begin{eqnarray}
\frac{\textrm{D}\rho}{\textrm{D}t}+\rho\nabla\cdot \textbf{v}=0,
\label{eq:cont}\\
\rho\frac{\textrm{D}\textbf{v}}{\textrm{D}t}=-\nabla
p+\frac{1}{c}\textbf{J}\times \textbf{B}, \label{eq:momen}
\\
\rho\frac{\textrm{D}}{\textrm{D}t}(e/\rho)=-p\nabla\cdot
\textbf{v},\label{eq:energy}\\
\frac{\partial \textbf{B}}{\partial t}=\nabla\times(\textbf{v}\times
\textbf{B}),\label{eq:induc}\\
\textbf{J}=\frac{c}{4\pi}\nabla\times \textbf{B},\label{eq:current}\\
p=(\gamma-1)e,\label{eq:p1}\\
p=\rho kT/m_{i}. \label{eq:p2}
\end{eqnarray}
Here $m_{i}$ is the proton mass, $\gamma$ is the ratio of specific heats, the
dependent variables are the magnetic field $\textbf{B}$,
the electric current density $\textbf{J}$, the mass density $\rho$,
the flow velocity $\textbf{v}$, the gas pressure $p$, and the internal
energy density $e$. We solve equations (\ref{eq:cont}) through (\ref{eq:p2}) numerically by using the
ZEUS-2D MHD code developed by Stone \& Norman (1992b).

The ZEUS-2D code is a two dimensional Eulerian finite difference code solving the equations of astrophysical
fluid dynamics including the effects of magnetic fields, radiation transport, self-gravity and rotation.
The code treats the divergence-free constraint ($\nabla\cdot{\bf B}=0$) with the constrained transport (CT) method of Evans $\&$ Hawley (1988), and uses the hybrid method of characteristics and constrained transport (MOC-CT) to maintain the stable and accurate propagation of all MHD wave modes. In the present work, radiation effects, self-gravity and rotation are neglected. More details of the code are available in the work by Stone \& Norman (1992a) and the follow-ups (Stone \& Norman 1992b; Stone et al. 1992).

We start with the magnetic configuration that consists of three components: a current-carrying flux rope that is used to model the prominence (or the filament) floating in the corona, the image of the current inside the flux rope, and the background magnetic field that is produced by a line dipole of the relative strength $M$ (see also Forbes 1990) located at the depth $d$ below the boundary surface $y=0$. The initial magnetic configuration from which the eruption occurs is given by (see also Forbes 1990)
\begin{eqnarray}
B_{x}&=&B_{\phi}(R_{-})(y-h_{0})/R_{-}-B_{\phi}(R_{+})(y+h_{0})/R_{+}\nonumber\\
&-&B_{\phi}(r+\Delta/2)Md(r+\Delta/2)[x^{2}-(y+d)^{2}]/R^{4}_{d},
\label{eq:Bx}\\
B_{y}&=&-B_{\phi}(R_{-})x/R_{-}+B_{\phi}(R_{+})x/R_{+}\nonumber\\
&-&B_{\phi}(r+\Delta/2)Md(r+\Delta/2)2x(y+d)/R^{4}_{d},\label{eq:By}
\end{eqnarray}
with
\begin{eqnarray*}
R^{2}_{\pm}&=&x^{2}+(y\pm h_{0})^{2},\\
R^{2}_{d}&=&x^{2}+(y+d)^{2}.
\end{eqnarray*}

The initial total pressure, including the gas pressure and the
magnetic pressure, as well as the mass density are
\begin{eqnarray}
p&=&p_{0}-\int^{\infty}_{R_{-}}B_{\phi}(R)j(R)d R,
\nonumber\\
\rho&=&\rho_{0}(p/p_{0})^{1/\gamma}, \label{eq:p_rho}
\end{eqnarray}
and $B_{\phi}(R)$ in equations (\ref{eq:Bx}) and (\ref{eq:By}) is
determined by the electric current density distribution $j(R)$
inside the flux rope, and reads as
\begin{eqnarray}
B_{\phi}(R)&=&-\frac{2\pi}{c}j_{0}R, \hspace{2mm} {\mbox for
}\hspace{2mm}0\leq R\leq r-\Delta/2,
\nonumber\\
B_{\phi}(R)&=&-\frac{2\pi j_{0}}{c R}\left\{\frac{1}{2} \left(r -
\frac{\Delta}{2}\right)^{2} - \left(\frac{\Delta}{\pi}\right)^{2} +
\frac{1}{2}R^{2} + \frac{\Delta R}{\pi}
\sin\left[\frac{\pi}{\Delta}\left(R-r+\frac{\Delta}{2}\right)\right]
\right.
\nonumber\\
&+& \left. \left(\frac{\Delta}{\pi}\right)^{2}
\cos\left[\frac{\pi}{\Delta}\left(R-r+\frac{\Delta}{2}\right)\right]\right\},
\hspace{2mm}{\mbox for }\hspace{2mm} r- \Delta/2 < R <r + \Delta/2,
\nonumber\\
B_{\phi}(R)&=&-\frac{2\pi j_{0}}{c
R}\left[r^{2}+(\Delta/2)^{2}-2(\Delta/\pi)^{2}\right],
\hspace{2mm}{\mbox for }\hspace{2mm} r+\Delta/2\leq R<\infty;
\nonumber\\
j(R)&=& j_{0}, \hspace{2mm}{\mbox for }\hspace{2mm} 0\leq R\leq
r-\Delta/2,
\nonumber\\
j(R)&=&\frac{j_{0}}{2}{\cos[\pi(R-r+\Delta/2)/\Delta]+1},\hspace{2mm}
{\mbox for } \hspace{2mm}r-\Delta/2<R<r+\Delta/2,
\nonumber\\
j(R)&=&0, \hspace{2mm}{\mbox for }\hspace{2mm} r+\Delta/2\leq
R<\infty.
\end{eqnarray}

\begin{figure}
\centering
\includegraphics[width=10cm,clip,angle=0]{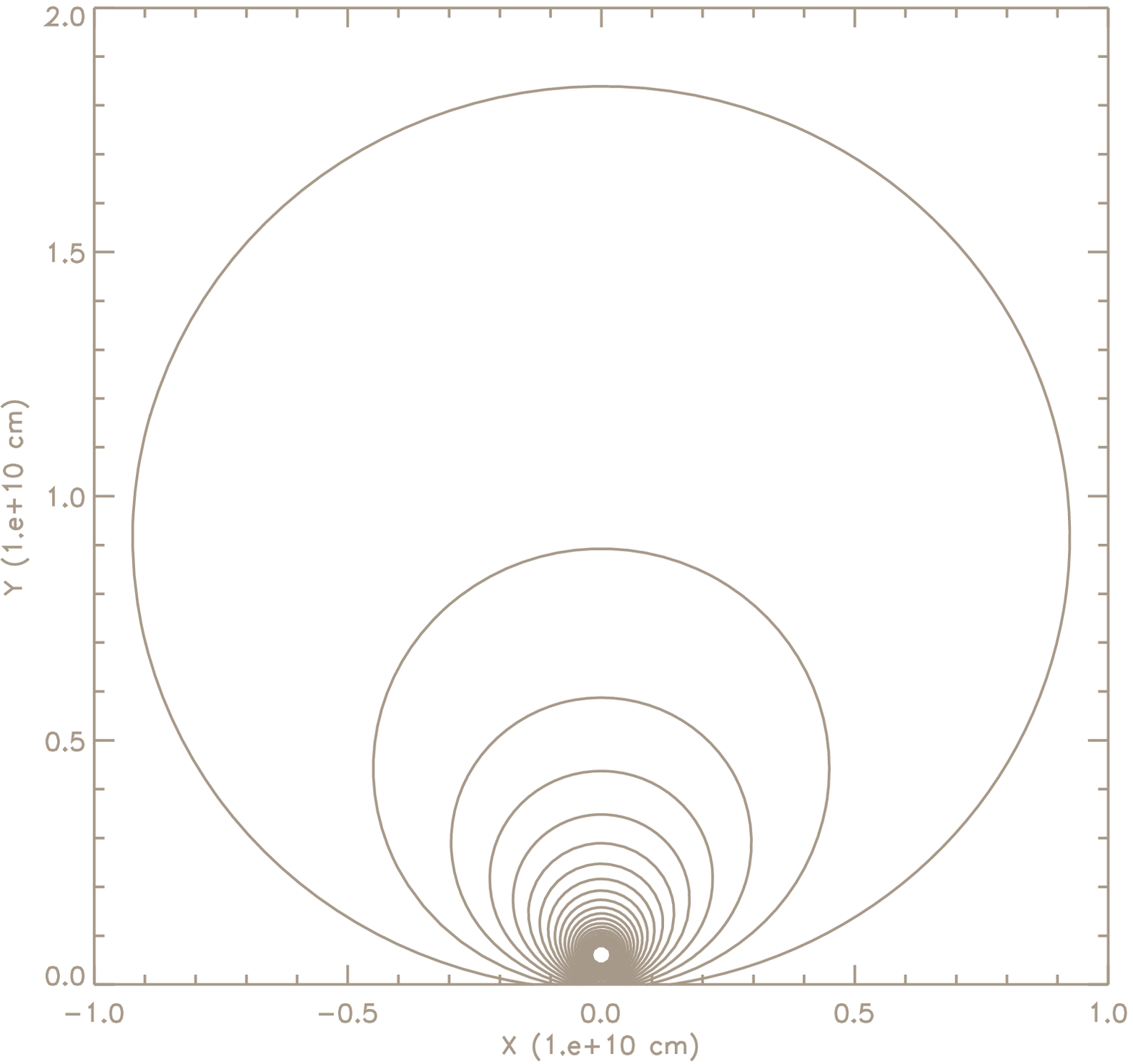}
\caption{The initial configuration of the magnetic field.}
\label{fig:t0}
\end{figure}

To this point, the code with the initial configuration (Figure \ref{fig:t0}) is ready for running with three important parameters fixed: the depth of the line dipole below the boundary surface: $d=3.125\times 10^{3}$ km, the relative strength of the line dipole $M=1$, and $\Delta=1.25\times 10^{3}$ km. The computational domain is taken to be $(-L, L)\times (0, 2L)$ with
$L=10^{5}$ km, and is discretized into $400\times 400$ grid points.
In the present work, we do not normalize parameters to the corresponding characteristic values as Forbes (1990) did. Instead we directly use the true values of these parameters required for running the code. When the system evolves, the radius $r$ and the height $h$ of the flux rope, and the electric current density inside the flux rope $j_{0}$ are functions of time; the plasma temperature $T$ and density $\rho$ depend on both time and the position in space. The initial values of these parameters are given in Table \ref{tbl:1}, and the ratio of the specific heats $\gamma=5/3$.

\begin{table}
\caption{Initial values for several important parameters of the
numerical experiment}
\begin{tabular}{lll}
\hline
$r_{0}=2.5\times 10^{3}$ km & $h_{0}=6.25\times 10^{3}$ km  & \\
$\rho_{0}=1.672\times10^{-12}$ g~cm$^{-3}$ & $T_{0}=10^{6}$ K & $j_{00}=1200$ statamp~cm$^{-2}$ \\
\end{tabular}
\label{tbl:1}
\end{table}

The value of $2.5\times 10^{3}$ km for $r$ gives a filament of just 5 times the grid resolution $\Delta x = 500$ km, so the internal structures of the filament is not well resolved. However, using such a small radius makes the box scale length much larger than the filament scale length, and this prevents the open boundary conditions from influencing the early evolution of the filament. The value of $j_{00}$ yields a strength of the magnetic field at the origin of about 200 G when the flux rope is absent.

In the present work, we are focusing on the wave-like phenomena caused by the disruption of an unstable magnetic configuration, so the initial values of those parameters given in Table \ref{tbl:1} are not necessarily for a configuration in equilibrium. The boundary condition at bottom side $y=0$ is a physical one, and those at other three sides are free boundary conditions.

Here, we note that $y=0$ in our previous works was either set up at the photospheric boundary (e.g., see discussions of Mei \& Lin 2008) or at the coronal base (e.g., see Lin \& van Ballegooijen 2005 and references therein) depending on the purpose of the specific work. In most cases, on the other hand, whether $y=0$ was set up at the photospheric boundary or at the coronal base does not matter because the distance between the two layers is small compared to the system lengthscale of interest (e.g., see Lin \& Forbes 2000, Lin et al. 2001, Lin et al. 2006). In the present work, however, $y=0$ is instead set up on the base of the chromosphere (or on the top of the photosphere) that is between the above two layers since one of the main purposes of this work is to investigate response of the chromosphere to the wave-like phenomena. Because the plasma in the photosphere is dense compared to that in the chromosphere and in the corona, the plasma density at and below $y=0$ can be taken as infinity, and magnetic field lines are line-tied to the boundary surface.

\section{Results of Numerical Experiments}
In this part of work, we present our results of the numerical experiments. Figure \ref{fig:t0} shows the initial configuration of the magnetic field. As we have mentioned shortly, this initial condition is not in equilibrium such that the magnetic compression surpasses the magnetic tension, so the flux rope starts to rise at the beginning of the experiment. With the lift-off of the flux rope, the closed magnetic field lines become stretched, and a reconnection region forms following the appearance of an X-type neutral point on the boundary surface. Diffusion by magnetic reconnection will convert magnetic energy to heating and the kinetic energy of the plasma.

We need to note here that no physical diffusion is included (see equation [\ref{eq:induc}]) in our experiment. So a current sheet should form when the magnetic field lines are severely stretched during the eruption as suggested by Forbes \& Isenberg (1991) and Lin \& Forbes (2000). In the numerical experiment, on the other hand, numerical diffusion substitutes the physical diffusion, and prevents the current sheet from forming and developing. Therefore, during the evolution in the system, there is no current sheet present in the magnetic configuration of interest. We do not invest much effort in removing the numerical diffusion in this work since our main goal is to study the MHD shocks and waves associated with CMEs, instead of the properties of the current sheet, or the reconnection region. Another reason why we leave this diffusion in the experiment is that magnetic reconnection invoked by it will help us look into several important details in the flare region.

\subsection{Formation and Propagation of the Shock around the Flux Rope}
With the experiment beginning, the flux rope starts to move from the height of $h_{0}=6.25\times 10^{3}$ km. At low altitudes and early phase, the propagation speed of the disturbance caused by the flux rope motion is less than local
magnetoacoustic speeds, so no shock forms until the disturbance travels faster than the magnetoacoustic wave.

Because of limited computational resources, we are unable to afford
to investigate the case that requires the grid resolution higher
than $400\times 400$. This is why we chose $\rho_{0} = 1.672\times
10^{-12}$~g~cm$^{-3}$, which is about two orders of magnitude larger
than that on the coronal base in reality ($\sim
10^{-14}$~g~cm$^{-3}$). But the strength of the magnetic field on
the base determined according to the parameters given in Table
\ref{tbl:1} is roughly the true value of magnetic strength in
reality. So, the Alfv\'{e}n speed in our calculations is about an
order of magnitude lower than that in reality (see Figures 1 and 2
of Lin 2002). A lower Alfv\'{e}n speed consequently allows the MHD
shock to form more easily. This may yield slower propagation and
weaker strength of various waves than in reality, but will not
change other properties of waves of interest as well as our
conclusions regarding the phenomena studied in this work. To further
simplify the problem, we focus in this section on the case in the
environment of the constant density, and will investigate the case
including the gravity in Section 3.3.

\begin{figure}
\centering
\includegraphics[width=14cm,clip,angle=0]{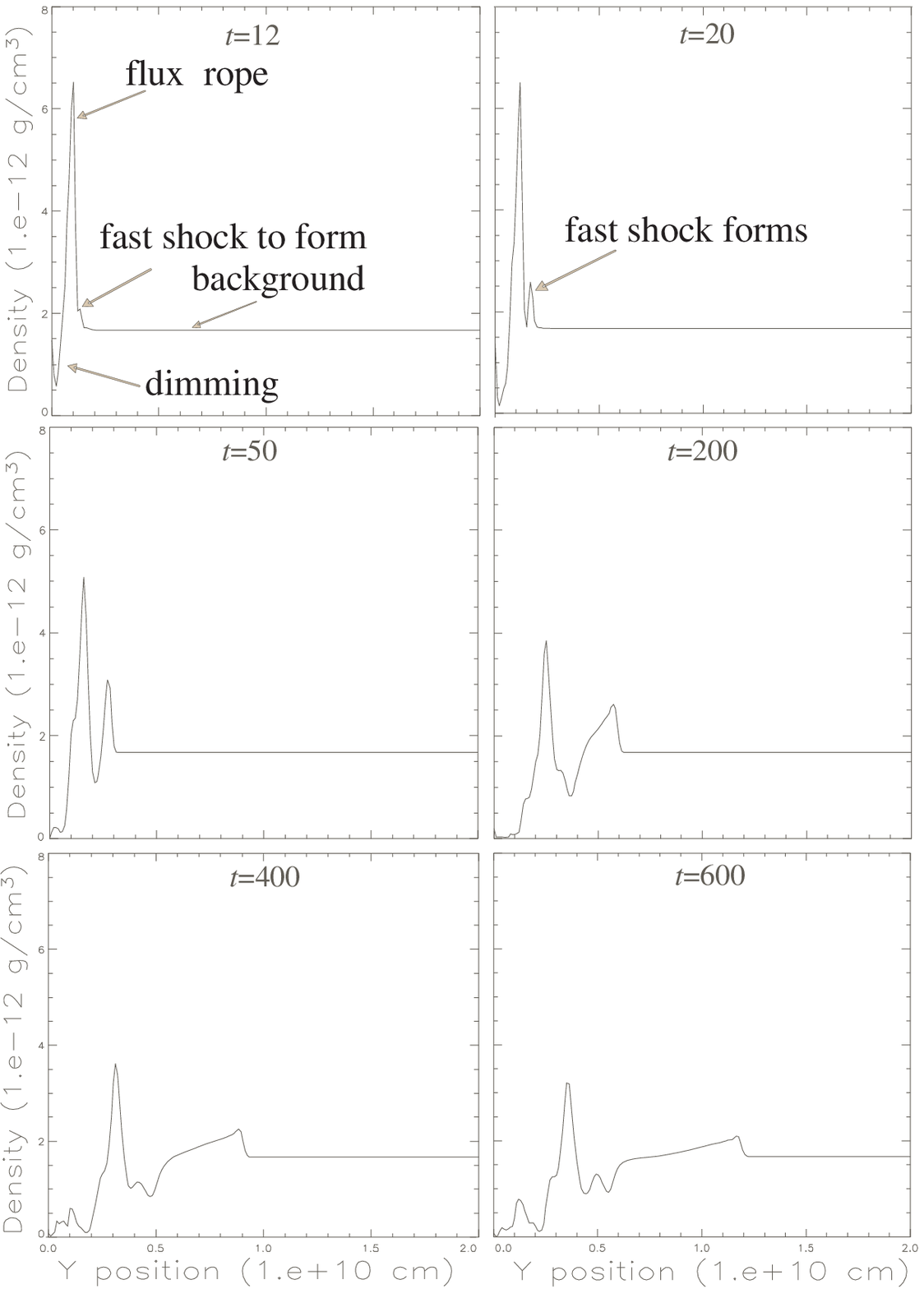}
\caption{The plasma density distribution along the $y$-axis at various times. The highest peak in each panel represents the flux rope that includes the densest material. In the panel of $t=12$~s, a slight increase in the density right forward of the flux rope begins to be recognized, which indicates the formation of the fast mode shock. A region behind the shock and surrounding the flux rope, in which the density is lower than the background density, corresponds to the dimming. The unit of time is second.}
\label{fig:x0d}
\end{figure}

The loss of equilibrium causes the flux rope to speed up, and a fast mode shock commences to form in front of it at around $t=12$ s. Figure \ref{fig:x0d} plots variations of the plasma density versus heights along the $y$-axis at different times. The highest peak in each panel represents the flux rope that includes densest material. In the panel of $t=12$~s, a slight increase in the density right forward of the flux rope begins to be recognized. The increase moves faster than the flux rope and gradually becomes significant. Because one of the most important features of the fast mode shock is the apparent enhancement of the plasma density, this second highest peak should thus represent the fast mode shock. In addition to the enhancement of the density caused by the shock, we also notice two dimming regions, one is right behind the fast mode shock and is shallow, and another is behind the flux rope at a farther distance and is deep. This scenario almost duplicates what Attrill et al. (2007) observed in a specific event (see Figure 3 of their paper), and should correspond to the dimming usually observed to be associated with CMEs (e.g., see Sterling \& Hudson 1997 and references therein). We shall discuss this issue in more details later.

To further confirm that the second peak shown in Figure \ref{fig:x0d} represents a shock, we are studying the Reynolds number of the fluid nearby the peak, $R_{e}=\delta y v/\nu$, where $\delta y$ is the full width of the peak at the half maximum, $v$ is its speed moving forward, and $\nu$ is the kinematic viscosity of the fluid. Basically, a shock develops from an ordinary wave with finite amplitude. In the propagation of the wave, the non-linear effect is important, and the crest of the wave moves faster than its leading or trailing edge. This causes a progressive steepening of the wave front portion as the crest catches up and, ultimately, the gradient of pressure, density, temperature and velocity become so large that dissipative processes, such as viscosity or thermal conduction, are no longer negligible. Then a steady wave-shape is attained with a balance between the steepening effect of the non-linear convective terms and the broadening effect of dissipation. Therefore, the Reynolds number of the fluid, $R_{e}$, around the shock should be of order of unity (e.g., see also detailed discussions given by Priest 1982, pp. 189--193).

\begin{figure}
\centering
\includegraphics[width=15cm,clip,angle=0]{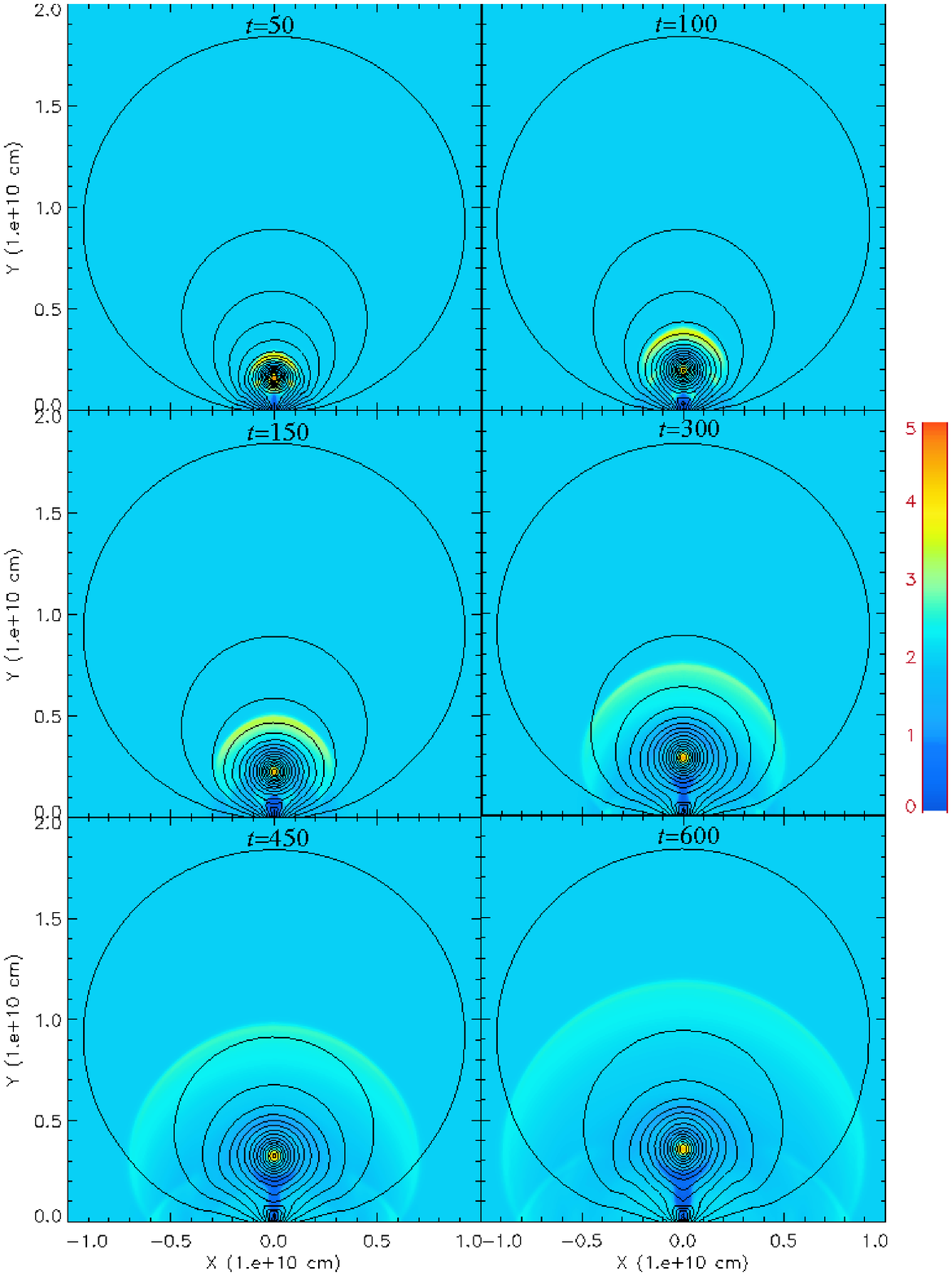}
\caption{Evolutions of the magnetic field and the plasma density as the eruption progresses. Continuous contours are magnetic field lines and the color shadings show the density distribution. Formation and development of the fast mode shock around the flux rope is clearly seen. The unit of time is second. The right color bar represents values of the density in $10^{-12}$~g~cm$^{-3}$.}
\label{fig:ad}
\end{figure}

In the present work, although we are dealing with a system that does
not include physical diffusive terms, the dissipation exists as a
result of the numerical viscosity. So the shock forms as the
numerical dissipation is balanced by the non-linear steepening
effect. Taking the case shown in Figure 2 at $t=50$~s as an example.
At the location where the second peak appears, the propagation speed
of the peak is $v=151$~km~s$^{-1}$, $\delta y = 10^{4}$~km, the
local Alfv\'{e}n speed is 37~km~s$^{-1}$, the sound speed is
117~km~s$^{-1}$, and the fast magnetoacoustic speed is
123~km~s$^{-1}$. In the original work of Stone \& Norman (1992a),
the numerical viscosity of the ZEUS-2D code was estimated, which
gave $\nu = 2.7 \times 10^{16}$~cm$^{2}$~s$^{-1}$. Bring values of
the above relevant parameters into the expression for $R_{e}$, we
get $R_{e} \approx 0.56$, which is roughly equal to unity. This,
together with comparison of values of various speeds, indicates that
the second peak shown in Figure \ref{fig:x0d} indeed represents the
fast mode shock.

Figure \ref{fig:ad} plots the evolutions of the magnetic field and the plasma density as the flux rope moves outward. Continuous contours (or curves) are magnetic field lines and the color shadings show the density distribution. As a denser area due to the enhancement in the density, the light yellow region in each panel corresponds to the fast mode shock. We see from these panels that the fast shock expands sidewards and backwards simultaneously as it propagates forward, and forms a crescent feature around the flux rope. At about $t=300$~s, it touches the boundary surface, and is then reflected by the boundary producing an echo at each of its footpoint, which propagates back into the corona.

This echo is a true phenomenon, not a numerical artefact. As approaching the bottom boundary, the shock enters a region where the plasma density changes in a dramatic way, and reflection and transmission of a wave (or shock) will take place in this region. Usually, rates of reflection and transmission are governed by the gradient of the density in this region. The larger the gradient is, the stronger the reflection is. In the case studied in this work, the density gradient could be infinity, so significant reflection of the wave (or shock) in this region is expected. Therefore, the echo shown in Figure \ref{fig:ad} is a physical result.

\begin{figure}
\centering
\includegraphics[width=15cm,clip,angle=0]{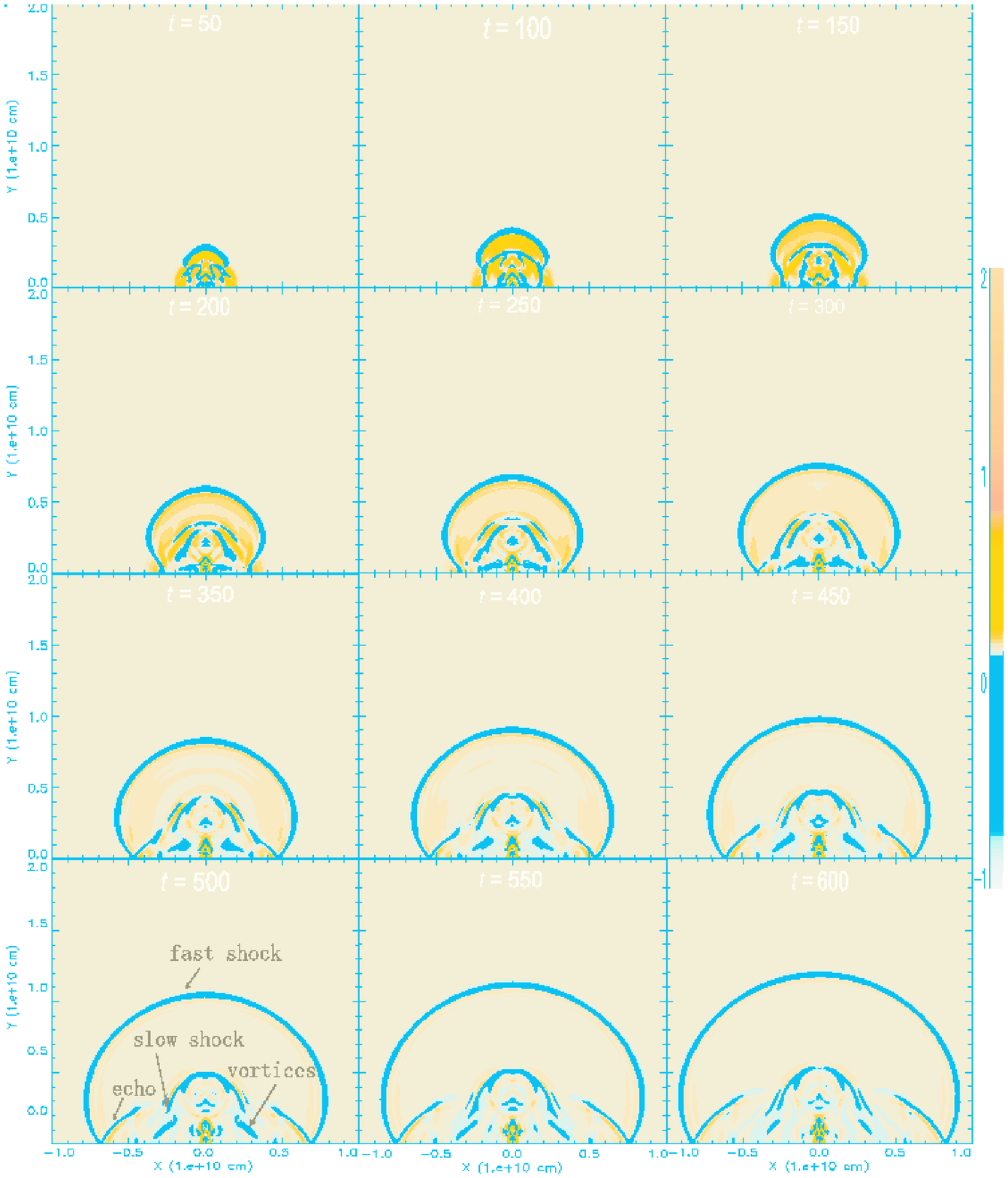}
\caption{Evolution of the velocity divergence. The fast shock, slow
shock, echo and vortices are denoted in the left bottom panel. The
unit of time is second. The right color bar represents values of the
velocity divergence in arbitrary unit.}
\label{fig:div}
\end{figure}

In the density contour plot, this echo is not apparent, so we study the velocity divergence $\nabla \cdot {\bf v}$ instead. Because the value of $\nabla \cdot {\bf v}$ is an indicator of the plasma compression, namely how much the plasma is compressed, changes in the value of $\nabla \cdot {\bf v}$ can well manifest the positions and propagations of the fast mode shock that causes the most significant compression of the fluid. Figure \ref{fig:div} shows the contours of $\nabla \cdot {\bf v}$ at various times as the disruption of magnetic field progresses. Compared to those in Figure \ref{fig:ad}, the characteristics of the fast mode shock in
Figure \ref{fig:div} are clear. Furthermore, formation of the echo after the shock touches the boundary surface can be easily recognized from the panel of $t=300$~s in Figure \ref{fig:div}, and propagations of the echo back into the corona are apparently shown in the subsequent panels as well. Both shock and echo are denoted in the left bottom panel of Figure \ref{fig:div}.

\begin{figure}
\centering
\includegraphics[width=10cm,clip,angle=0]{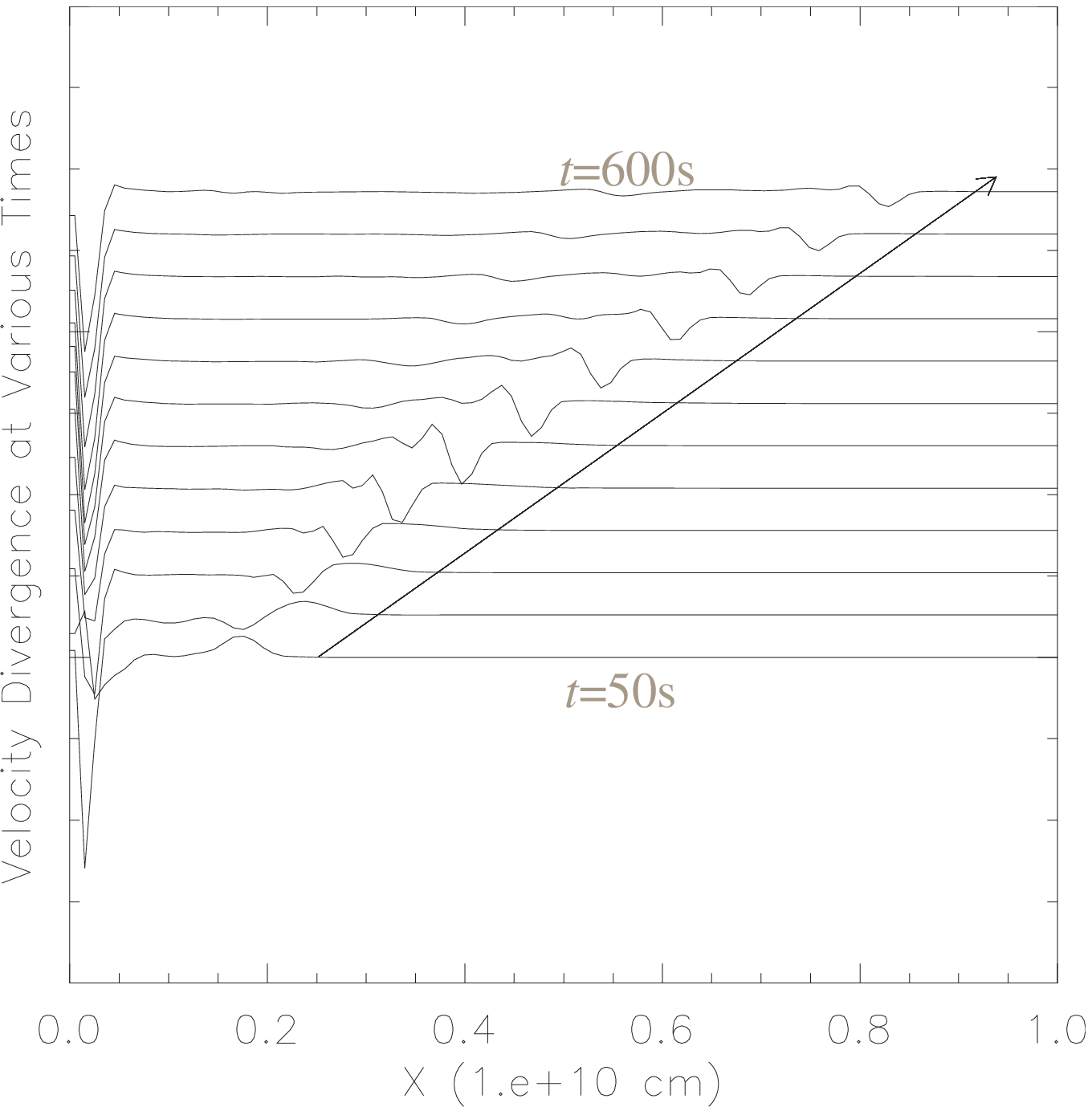}
\caption{Evolution of $\nabla\cdot{\bf v}$ distribution on the bottom boundary $y=0$ in the time interval between 50~s and 600~s. Changes in the shape of each curve from time to time apparently suggest a moving object sweeping the boundary layer.}
\label{fig:y0d}
\end{figure}

Obviously, the location where the fast shock touches the boundary surface moves outwards associating with the shock propagation. To reveal more details of this process, we study the temporal behavior of the distribution of $\nabla\cdot {\bf v}$ in space on $y=0$. We plotted 12 curves in Figure \ref{fig:y0d} with each one for the distribution of $\nabla\cdot {\bf v}$ on the boundary $y=0$ from
$x=0$ to $x=10^{5}$ km at a specific time. The curve at the bottom corresponds to the time $t=50$ s, the one at the top to $t=600$ s, and the time interval between every two adjacent curves is 50 s. Most part of each of these curves is a flat line. This is due to the limited propagation speed of the disturbance, and plasma at large distance is not disturbed until the footprints the shock reaches. Therefore, the sequences of the moving features on each curve (see also the arrow in Figure \ref{fig:y0d}) actually represents the propagation of the response of the boundary surface to the shock wave, which is implicitly suggestive of the observable moving footprints of the fast mode shock in the chromosphere during the eruption. Further calculation gives a speed of 126~km~s$^{-1}$ for the propagation. We account such a propagating feature for the Moreton wave observed in H$\alpha$. Consulting Figure 2 of Lin (2002), we find that, in this work, the Alfv\'{e}n speed in the area around $y=0$ where the Moreton wave supposes to appear is less than 50~km~s$^{-1}$, which confirms the super-Alfv\'{e}nic property of the Moreton wave.

\begin{figure}
\centering
\includegraphics[width=15cm,clip,angle=0]{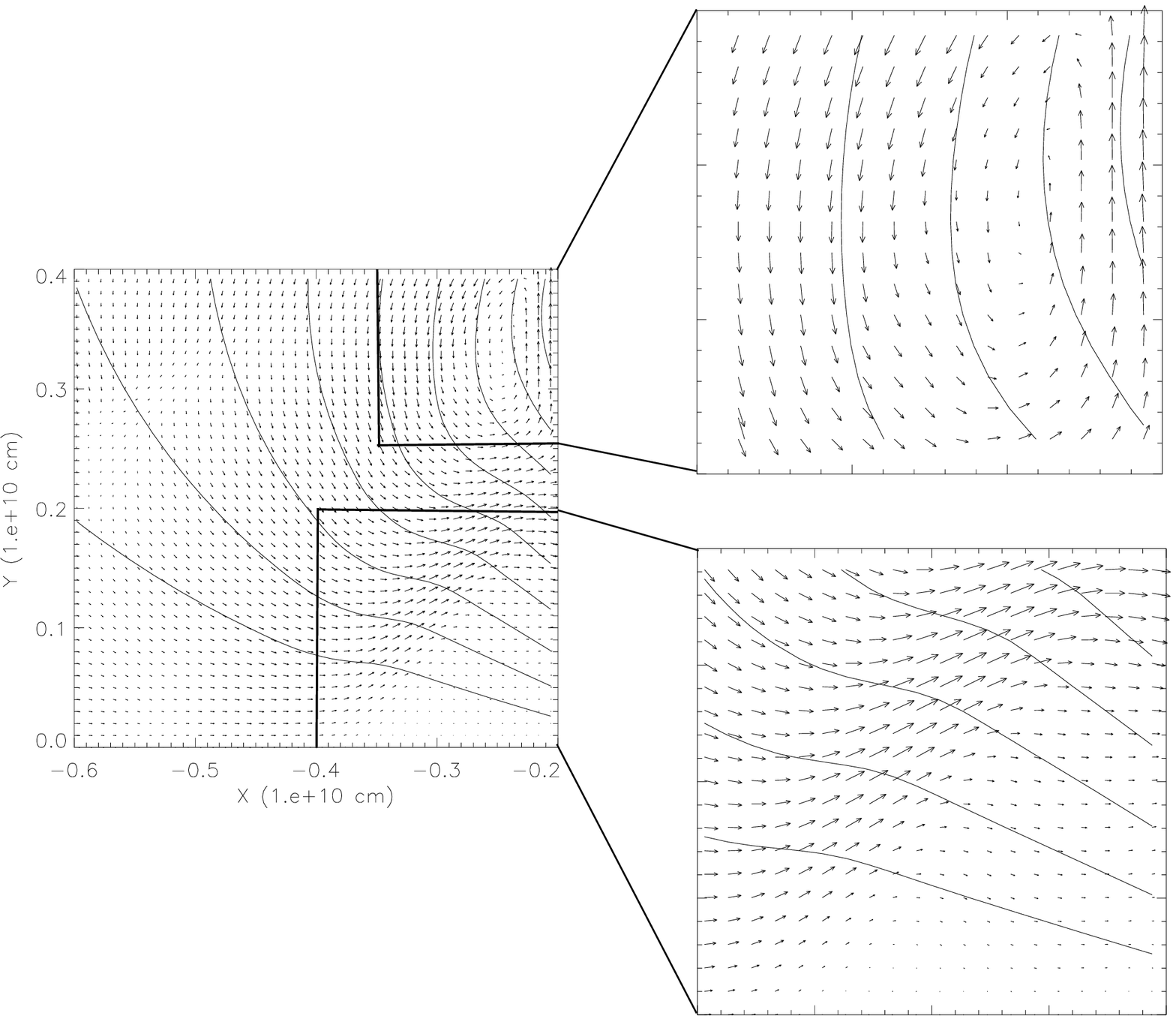}
\caption{Magnetic field lines and streamlines at $t=600$~s. The left
panel provides a largescale view of both field and stream lines. Two
regions surrounded by rectangles display areas in which the flow is
strong and the magnetic field lines are apparently deformed. Details
in these two regions can be seen more clearly in the enlarged panels
on the right.}
\label{fig:vortcs}
\end{figure}

In addition to the fast mode shock, we are also able to recognize several other features, including the slow mode shock developing from two sides of the flux rope and the velocity vortices on either side of the reconnection region. All of them are denoted in the left bottom panel in Figure \ref{fig:div}. To look into more details of the velocity vortices and their impact on the nearby magnetic field, we enlarged a local region of the right bottom panel in Figure \ref{fig:ad} and create a composite of magnetic field lines and streamlines as shown in Figure \ref{fig:vortcs}. The left panel in Figure \ref{fig:vortcs} provides a largescale view of both field and stream lines. Two regions surrounded by rectangles display areas in which the flow is strong and the magnetic field lines are apparently deformed. Details in these two regions can be seen more clearly after they are further enlarged as shown on the right. The plasma flow manifests apparent vortices near the side back of the flux rope (see the upper panel). The vortices behave the same way as noticed by Forbes (1990) such that they do not persist for a very long time at any given location. Instead they propagate outward toward the open boundaries. Unlike the fast mode shock, the vortices could not reach the boundary surface at $y=0$.

Besides the velocity vortices, the plasma flow also show the motion toward the reconnection region (see the lower panel in Figure \ref{fig:vortcs}). As a result of the disruption of the magnetic configuration, the flux rope moves rapidly upward and leaves a region of lower pressure behind. This yields two flows toward the lower pressure region from opposite direction bringing in both plasma and magnetic fields of opposite polarity. Forcing magnetic fields of opposite to move together consequently results in the driven magnetic reconnection in that region. Although the MHD equations (\ref{eq:cont}) through (\ref{eq:p2}) governing the system do not include the physical diffusion term, the numerical diffusion as we have discussed plays the role in dissipating the magnetic field (e.g., see also discussions of Chen et al. 2009).

Comparison among various speeds suggests the formation of the fast shock in front of the flux rope. The slow magnetoacoustic speed is smaller than the fast one, so there should be the slow mode shock present somewhere associated with the fast shock formation. Because the slow shock moves apparently slower than the fast shock, the magnetic field behind the slow shock is refracted toward the normal of the shock and its strength decreases as the shock front passes by, the slow shock is not very far from the flux rope working as a piston driving both shocks. Comparing the features of magnetic field and the plasma flow around the flux rope, we are able to identify a pair of slow shocks in the flank of the flux rope as indicated in Figure \ref{fig:div}. After identifying the slow shocks, we further notice that the slow shock also expands backward to the bottom boundary like the fast one. But like the velocity vortices, the slow shock cannot reach the boundary, either. It decays and disappears somewhere above the boundary surface.

In the region where the slow shock contacts the vortices, the $\nabla \cdot {\bf v}$ contours show complex features. Formation of the velocity vortices are also indicated by the deformation of the magnetic field lines around the reconnection region (see the bottom two panels in Figure \ref{fig:ad} and those in Figure \ref{fig:vortcs}). The fact that the slow mode shock and the velocity vortices does not reach the boundary surface can also be seen from the plots in Figure \ref{fig:y0d} that show the signs of the fast mode shock only.

\begin{figure}
\centering
\includegraphics[width=15cm,clip,angle=0]{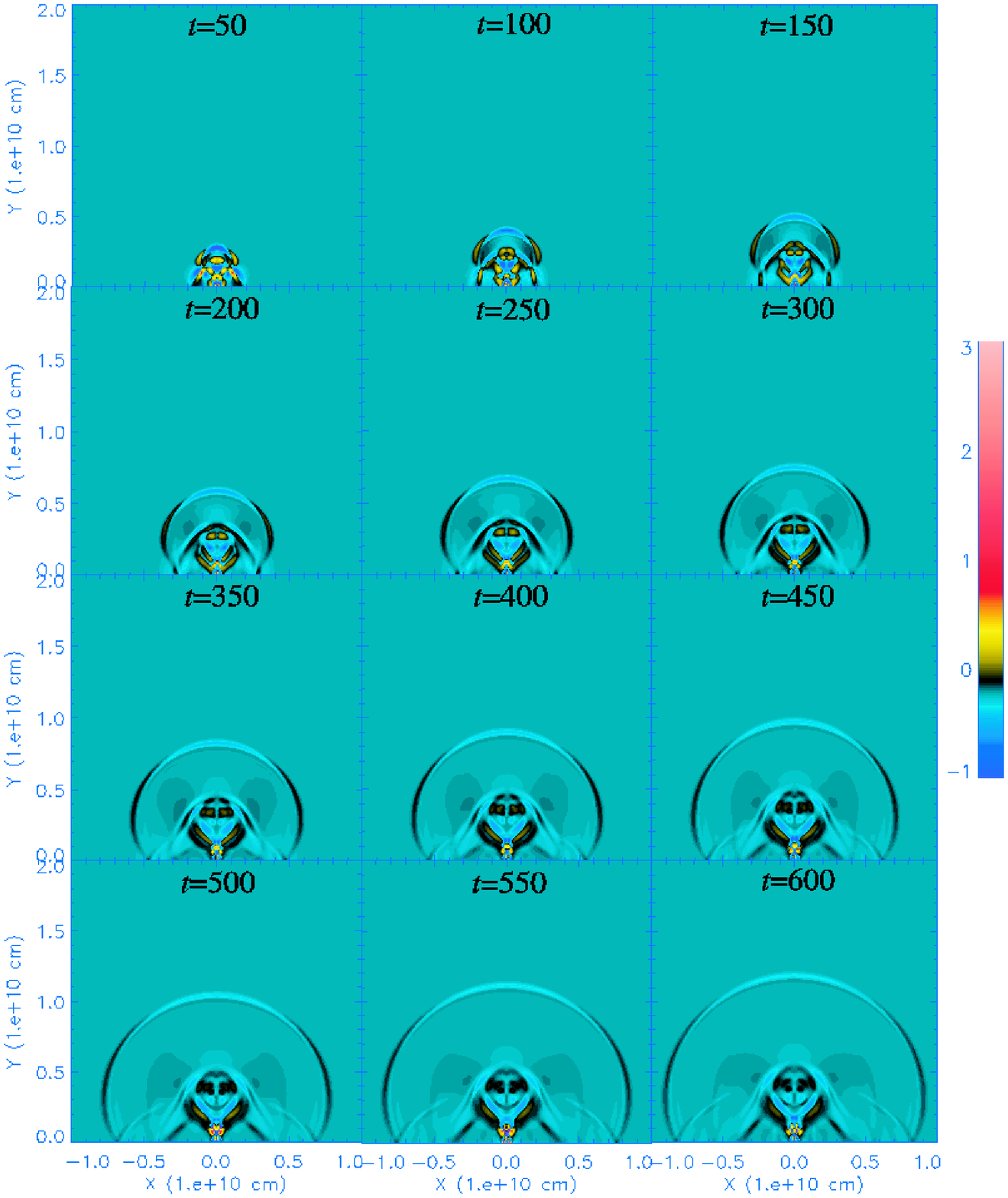}
\caption{A set of snapshots of $\nabla\times{\bf v}$ in the time interval between 50 s and 600 s. In addition to those features shown in Figure \ref{fig:div}, an extra characteristic that can be recognized here is a pair of the Petschek slow mode shocks developing from the reconnection region. The unit of time is second. The right color bar represents values of the velocity divergence in arbitrary unit.}
\label{fig:vor}
\end{figure}

Results of Forbes (1990) showed that the velocity vorticity $\nabla\times{\bf v}$ is a good tracer of the slow mode shock, especially that developed by the Petschek (1964) reconnection. According to Forbes (2009, private communications), this has to do with the relative size of the velocity changes across the shock. In the fast mode shock, the normal velocity ($\nabla\cdot {\bf v}$ mainly results from its change) can at most change by a factor of 4 (e.g., from 1 upstream to 0.25 downstream). Typically the change is smaller since factor 4 can only be reached when the magnetic field vanishes. In the case of slow mode shock, the normal velocity across a slow shock in the switch off limit can at most change by a factor of 5/3 (the plasma $\beta=0$ limit). However, it is the tangential velocity that is important for determining the vorticity. For the slow mode shock near the switch off limit, the tangential velocity is capable of undergoing an enormous change. It can be nearly the ambient Alfv\'{e}n speed upstream, and then decreases to zero downstream. Therefore, the velocity vorticity $\nabla\times{\bf v}$ should reveals more information on the slow shock.

In two-dimensional configurations as we study here, vector of the velocity vorticity $\nabla\times {\bf v}$ has the $z$-component only. Figure \ref{fig:vor} shows a set of snapshots of $(\nabla\times{\bf v})_{z}$ in the time interval between 50 s to 600 s corresponding to those in Figures \ref{fig:ad} and \ref{fig:div}. In each panel, the fast shock and its echoes can still be seen clearly, but the slow shock becomes more apparent. Besides the slow shocks that we already mentioned above, another two pairs of the slow mode shocks as a result of the Petschek-type reconnection process (e.g., see also Petschek 1964, Priest 1982, and Priest \& Forbes 2000) are recognized to form from an X-type neutral point below the flux rope.

\begin{figure}
\centering
\includegraphics[width=17cm,clip,angle=0]{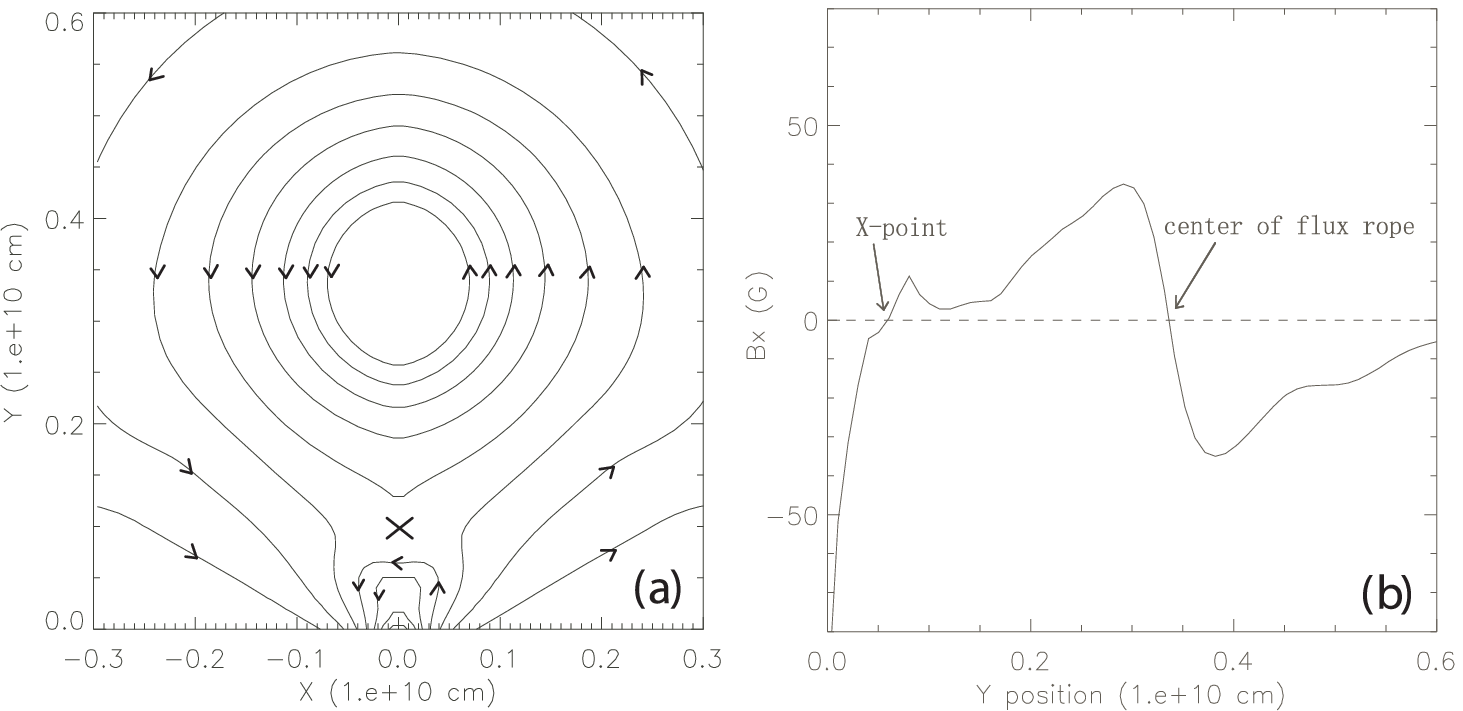}
\caption{The local magnetic field that includes both the flux rope and the X-point (marked by $\times$) at $t=500$ s (a), and the corresponding distribution of $B_{x}$ on the $y$-axis (b). Positions of the X-point and the center of the flux rope (namely the O-point) are denoted in (b).}
\label{fig:mc500}
\end{figure}

As demonstrated in previous works (e.g., see Forbes \& Isenberg
1991; Isenberg et al. 1993; Forbes \& Priest 1995; Lin \& Forbes
2000; Lin et al. 2006), magnetic dissipation in the magnetic
configuration of interest undergoes very slowly or is almost
forbidden due to the high electric conductivity in the corona
environment unless a neutral point or a current sheet appears in the
configuration. The loss of equilibrium in the system stretches the
closed magnetic field such that a neutral point or a current sheet
forms between the boundary surface and the flux rope. It is this
X-point or current sheet that allows fast magnetic dissipation in
the form of reconnection to occur, converting magnetic energy into
heating and kinetic energy at a reasonably rapid rate. Figure
\ref{fig:mc500}a plots the magnetic field contours for the region
below the flux rope. In this region an X-type neutral point exists.
Figure \ref{fig:mc500}b plots the distribution of magnetic field
$B_{x}$ along the $y$-axis to further confirm that an X-point does
exist in that region. Due to the symmetry of the configuration,
$B_{y}$ vanishes on the $y$-axis. So investigating the behavior of
$B_{x}$ on the $y$-axis alone should be able to provide us the
information needed for demonstrating the existence of the X-point.

Two pairs of slow mode shocks develop from the X-point. These shocks are similar to those proposed originally by Petscheck (1964), except that the downward ones are not very apparent since they usually dissociate into isothermal shocks and conduction fronts (see discussions of Forbes \& Acton 1996 for more details). But those propagating upward remain recognizable during the whole process. Associated with this pair of slow mode shocks is the reconnected plasma and magnetic flux that enters the CME bubble, increasing the amount of mass and magnetic flux in the CME (or interplanetary CME) that eventually flows into interplanetary space as suggested by Lin et al. (2004).

With the evolution progressing, the slow shocks invoked around the flux rope extend side-back to the flank of the flux rope as well, similar to the fast one. As shown by those in Figure \ref{fig:div}, panels in Figure \ref{fig:vor} clearly indicate that the layers in the lower atmosphere where the fast and the slow shocks can reach are quite different. To further address this point, we study the distribution of $(\nabla\times{\bf v})_{z}$ on the boundary surface $y=0$ as we
did for Figure \ref{fig:y0d}.

\begin{figure}
\centering
\includegraphics[width=18cm,clip,angle=0]{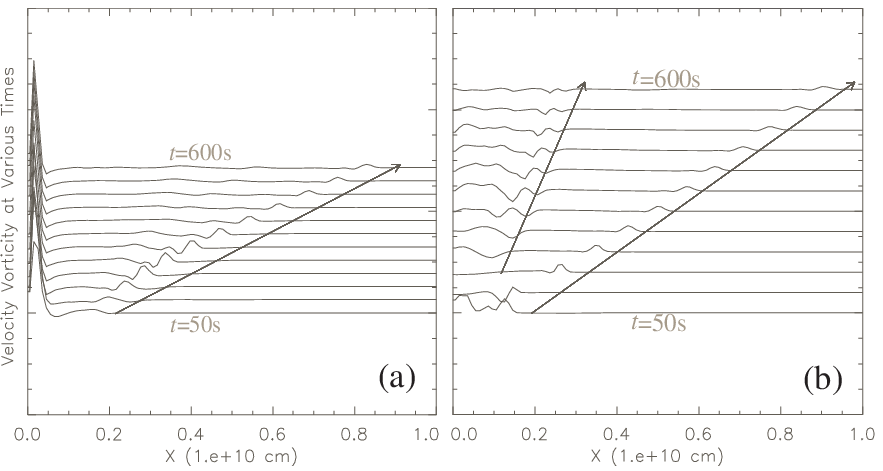}
\caption{Distributions of the $z$-component of the velocity vorticity
$\nabla\times{\bf v}$ on the boundary layer $y=0.0$ (a), and on the layer of $y=0.3$ (b) in the time interval between 50~s and 600~s.}
\label{fig:yv}
\end{figure}

Figure \ref{fig:yv}a plots 12 $(\nabla\times{\bf v})_{z}$ curves
versus $x$ on $y=0$ in the time interval between $t=50$~s and
$t=600$~s. Like what we see from Figure \ref{fig:y0d}, plots in
Figure \ref{fig:yv}a show clear sign of the fast shock that sweeps
the boundary surface at speed of around 126~km~s$^{-1}$. But these
plots do not display any sign of the slow mode shock or the velocity
vortices. Figure \ref{fig:vor} suggests that the disturbance caused
by the slow shock and the velocity vortices should exist in the
higher layers. Testing indicates that the impact of the slow shock
starts appearing above the layer of $y=0.3$.

As an example, we duplicate the 12 plots shown in Figure \ref{fig:yv}a for $y=0.3$, and the results are displayed in Figure \ref{fig:yv}b. Difference between Figures \ref{fig:yv}a and \ref{fig:yv}b is significant. The plots manifest clearly two propagating features at layer $y=0.3$, one moves faster and the other moves slower. Obviously, the faster one is the disturbance caused by the fast shock in the corona, and it is quite likely to account for the X-ray waves reported by Rust \& \v{S}vestka (1970), Khan \& Aurass (2002), and Narukage et al. (2004). The slower one should be the footprint of either the slow shock or the velocity vortices or the both in the corona. According to the distance it propagates at specific times, we find that the slow shock sweeps the coronal layer of $y=0.3$ at speed of about 50~km~s$^{-1}$, which is around 40\% the speed the fast shock sweeps the boundary layer $y=0.0$. Considering the observational features of the EIT wave and the layer where the EIT wave is seen in the solar atmosphere
(e.g., see Thompson et al. 1998; Atrill et al. 2007; Wills-Davey et al. 2007; Thompson \& Myers 2009), we conclude that these results are highly suggestive of EIT waves originating from the disturbance by the slow mode shocks and the velocity vortices in the corona.

\subsection{Formation of Dimming Areas and Disturbance Nearby}
The dimming, which we are discussing here and is commonly mentioned
and studied in the community, is the region observed darker than the surrounding area in the eruption. It is believed that the magnetic field configuration is severely stretched by the eruption in a transient way, causing the volume of the magnetic structure to increase and plasma density density in the structure to decrease quickly, strongly depletes the coronal emission, and forms dimming regions on the solar disk (e.g., see Forbes \& Lin 2000; Harrison \& Lyons 2000; Harrison 1997; Zarro et al. 1999; Thompson et al. 2000; Harrison et al. 2003). Therefore, the dimming, we are studying here, is not caused by the low temperature plasma from the chromosphere flowing into the coronal structure. Instead it is caused by the density decrease as a result of the eruption.

\begin{figure}
\centering
\includegraphics[width=14cm,clip,angle=0]{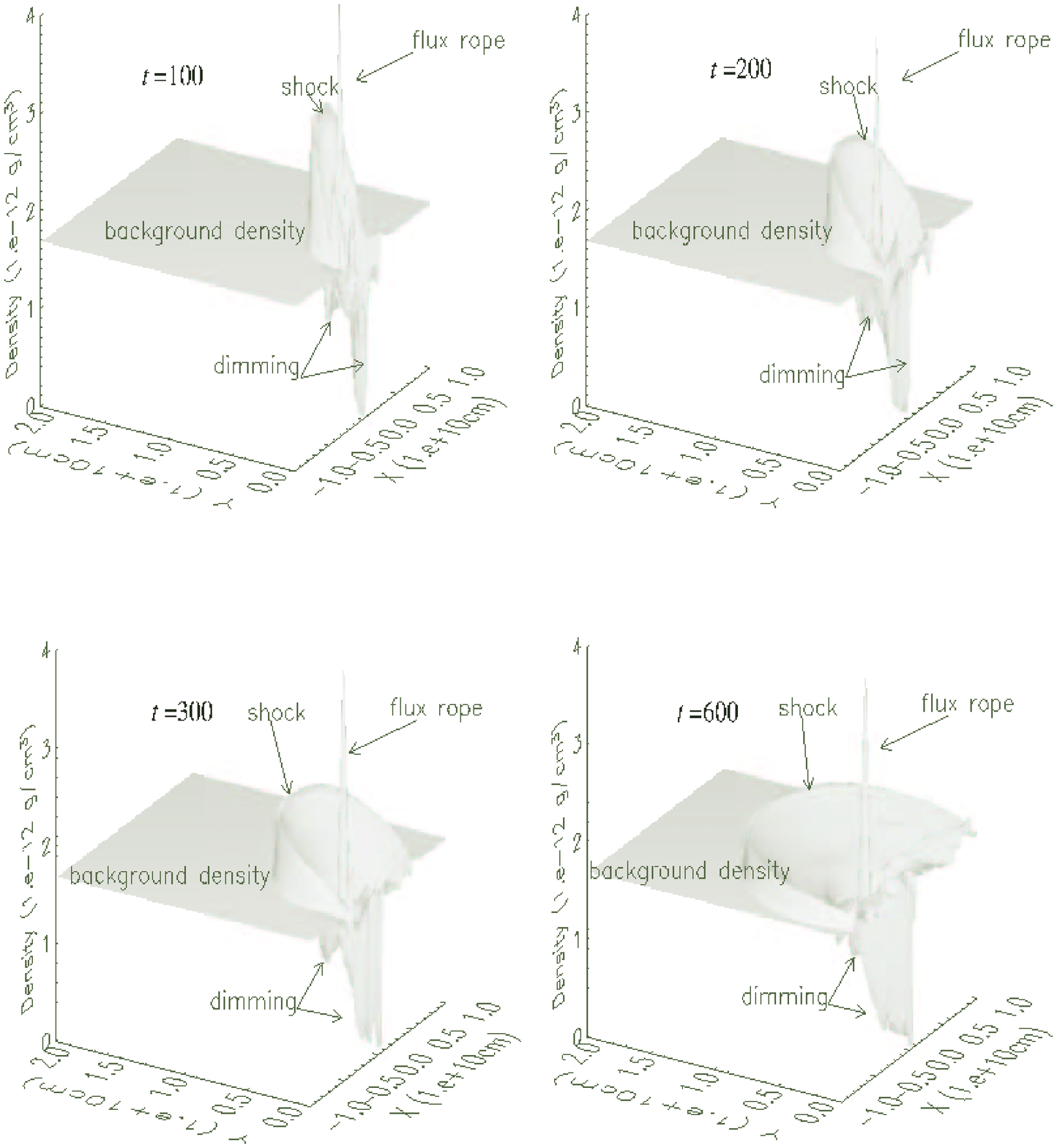}
\caption{Density distributions on the $x$-$y$ plane at various times. The arrows specify the flux rope, the fast mode shock and the dimming regions, respectively. The plane in the middle indicates the background density $\rho=1.67\times10^{-12}$~g~cm$^{-3}$. The unit of time is second.}
\label{fig:d}
\end{figure}

The low density region that forms in our numerical experiment can be seen in Figure \ref{fig:ad}. It is shown in blue color. This region keeps expanding as the eruption progresses. Although Figure \ref{fig:x0d} provided some information of the dimming in the eruption, it is only for that along the $y$-axis. To display evolutionary behaviors of this whole region, we plot in Figure \ref{fig:d} variations of plasma density in the $x$-$y$ plane versus time. In each panel, the $x$-$y$ plane extends horizontally in the direction perpendicular to the paper plane, and the direction perpendicular to this plane shows the change in the density. The plane in the middle located at $\rho=1.67\times10^{-12}$ g cm$^{-3}$ indicates the background density, in addition, the highest peak (i.e., the high mass density) represents the flux rope, the crescent feature represents the fast mode shock, and the area that the density is below the background illustrates the dimming. These features have been specified in each panel in Figure \ref{fig:d}. In the region between the flux rope and the fast mode shock, we noticed there is the disturbance in density present as well. Comparing with Figure \ref{fig:x0d}, Figure \ref{fig:d} apparently provides more details of the dimming region, and well duplicates the observational results of Attrill et al. (2007).

Before ending this part of work, we need to address one more point regarding the type II radio burst and its origin in eruption. Whether the type II burst originates from the CME-driven shock or from the flare-blast wave has long be an open question (e.g., see Lin et al. 2006 and references therein), and is still a subject of active researches.

Whether a flare could ignite a type II radio burst actually depends
how we define a flare. If we follow the traditional definition
(e.g., see \v{S}vestka 1976) or the term, the eruptive solar flare,
given by Forbes \& Isenberg (1991), then any eruptive process could
be classified into the category of flare that may produce the burst.
With the development of observations and theoretical studies, on the
other hand, the definition of various manifestations of an eruptive
process becomes more specific. For example, we usually distinguish
solar flares, eruptive prominences, and CMEs from on another; and
compared to the CME, the flare occupies a small volume that includes
flare ribbons and loops only.

In this work, we follow the second definition. So the speed of the fastest motion of the large scale structure in the flare region is less than 100 km/s in reality (corresponding to 10 km/s in our calculations), which is the speed of flare loop and ribbon motions (\v{S}vestka 1996; Lin 2002). As many studies and investigations have shown that, the motions of the flare loops and ribbons are not involved in any mass motion (e.g., see Schmieder et al. 1987; Forbes \& Acton 1996). Instead these motions result from the successive propagation of the energy release site (or reconnection site) from old field lines to the new ones. Therefore, in any sense, it is hard to find an object in the flare region moving fast enough to ignite the so-called blast wave that could account for the type II radio burst (e.g., see also discussions of Lin et al. 2006). Furthermore, the difference in the plasma densities in the flare region (10$^{12}$ cm$^{-3}$) and at the positions where type II radio bursts usually start to be observed (10$^{6}$ cm$^{-3}$) also implies that the type II radio burst cannot be produced in the flare region.

\begin{figure}
\centering
\includegraphics[width=10cm,clip,angle=0]{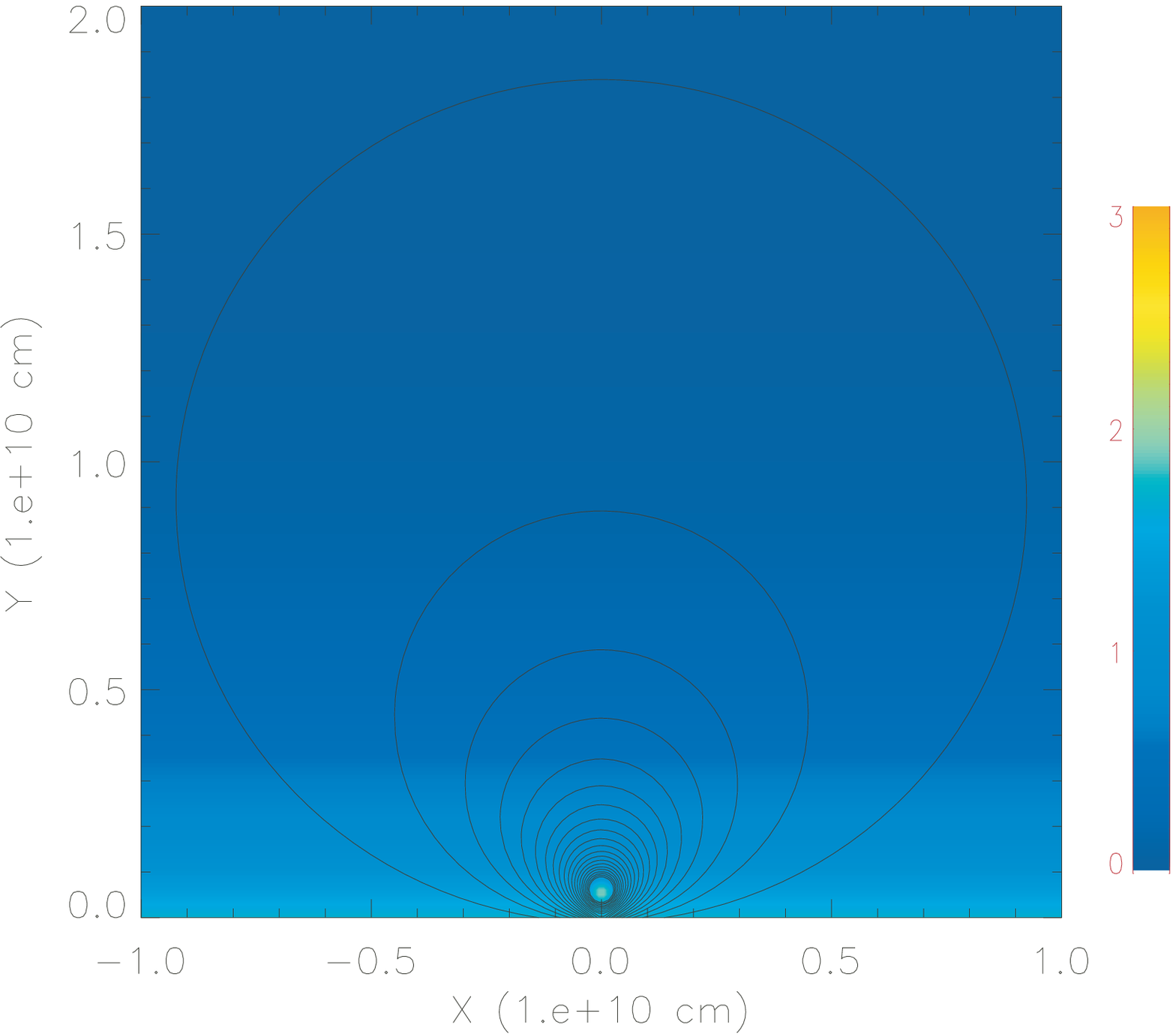}
\caption{The initial configuration of the magnetic field and the plasma density distribution in the environment including the gravity. Continuous contours are magnetic field lines and the color shadings show the density distribution. Stratification of the atmosphere as a result of the gravity is seen clearly. The right color bar represents values of the density in $10^{-12}$~g~cm$^{-3}$.} \label{fig:gart0}
\end{figure}

Thus the flare in such a scenario cannot be a driver of the blast wave that is able to account for the type II radio burst although the relevant debate on this issue may last for a while. In fact, evolutionary features of the disrupting magnetic configuration shown in Figures \ref{fig:ad}, \ref{fig:div}, \ref{fig:vortcs}, and \ref{fig:vor} also suggest the absence of any energetic wave-like phenomenon around the flare region in addition to the slow mode shocks and velocity vortices. Our results indicate that the only igniter that is capable of invoking the burst is the fast mode shock driven by the CME (or the flux rope). This confirms the conclusion of Lin et al. (2006) that there is no agent in the flare region that could be responsible for the ignition of the type II radio burst, that the so-called flare-blast wave, in fact, does not exist, and that the CME-driven fast shock is the sole source of the burst, at least our results indicate so.

\subsection{Evolution in the Environment Including the Gravity}
In this section, with the consideration of the gravity, we re-perform the work that has been done shortly. We use the same numerical approaches and formulae except that equation (2) now includes the gravity, $\rho GM_{\odot}/(R_{\odot}+y)^{2}$. Here the gravitational constant $G=6.672\times
10^{-8}$~dyn~cm$^{2}$~g$^{-2}$, the solar mass
$M_{\odot}=1.989\times 10^{33}$~g, and the solar radius
$R_{\odot}=6.963\times10^{10}$~cm. Equations in (\ref{eq:p_rho}) now read as
\begin{eqnarray}
p&=&n_{0}kT_{0}\exp\left[\frac{GM_{\odot}m_{p}}{kT_{0}}\left(\frac{1}{R_{\odot}+y}-\frac{1}{R_{\odot}}\right)\right]
-\int^{\infty}_{R_{-}}B_{\phi}(R)j(R)dR, \nonumber\\
\rho&=&n_{0}m_{p} (p/p_{0})^{1/\gamma}, \label{eq:p_rho_g}
\end{eqnarray}
where
\begin{eqnarray*}
p_{0}=n_{0}kT_{0}\exp\left[\frac{GM_{\odot}m_{p}}{kT_{0}}\left(\frac{1}{R_{\odot}+y}-\frac{1}{R_{\odot}}\right)\right],
\end{eqnarray*}
$k$ is the Boltzmann constant, and the plasma density on the bottom boundary $n_{0}=10^{12}$~cm$^{-3}$ (or $\rho_{0}=1.672\times 10^{-12}$ g~cm$^{-3}$). Note that we are still working on the isothermal case here for two reasons, one is the the simplicity of mathematics, and another one is because the true corona is roughly isothermal at lower altitude, say within the range of around 0.7 $R_{\odot}$ (see Figure 1 of Lin 2002 as well as the result of Sittler \& Guhathakurta 1999).

Figure \ref{fig:gart0} shows the initial configuration of the magnetic field and the initial density distribution with the gravity included. Continuous contours are magnetic field lines and colorful shadings show the density distribution. We notice that the plasma density decreases with height $y$. The parameters for the initial magnetic configuration remain unchanged. So the initial structure is not necessarily in equilibrium. Following the practice performed in previous sections, we start evolving the system from the non-equilibrium state.

In the isothermal environment including the gravity, the plasma
density decreases with altitude roughly in the exponential way as
suggested by (\ref{eq:p_rho_g}). In this case, the Alfv\'{e}n speed
may become large at higher altitude (e.g., see Figures 1 and 2 of
Lin 2002), and magnetic reconnection is allowed to take place at a
reasonably high rate (e.g., see also discussions of Lin \& Forbes
2000). High rate of reconnection means fast energy conversion, and
fast motion of the flux rope. So the flux rope motion should be
faster here than in the case of constant plasma density.

\begin{figure}
\centering
\includegraphics[width=14cm,clip,angle=0]{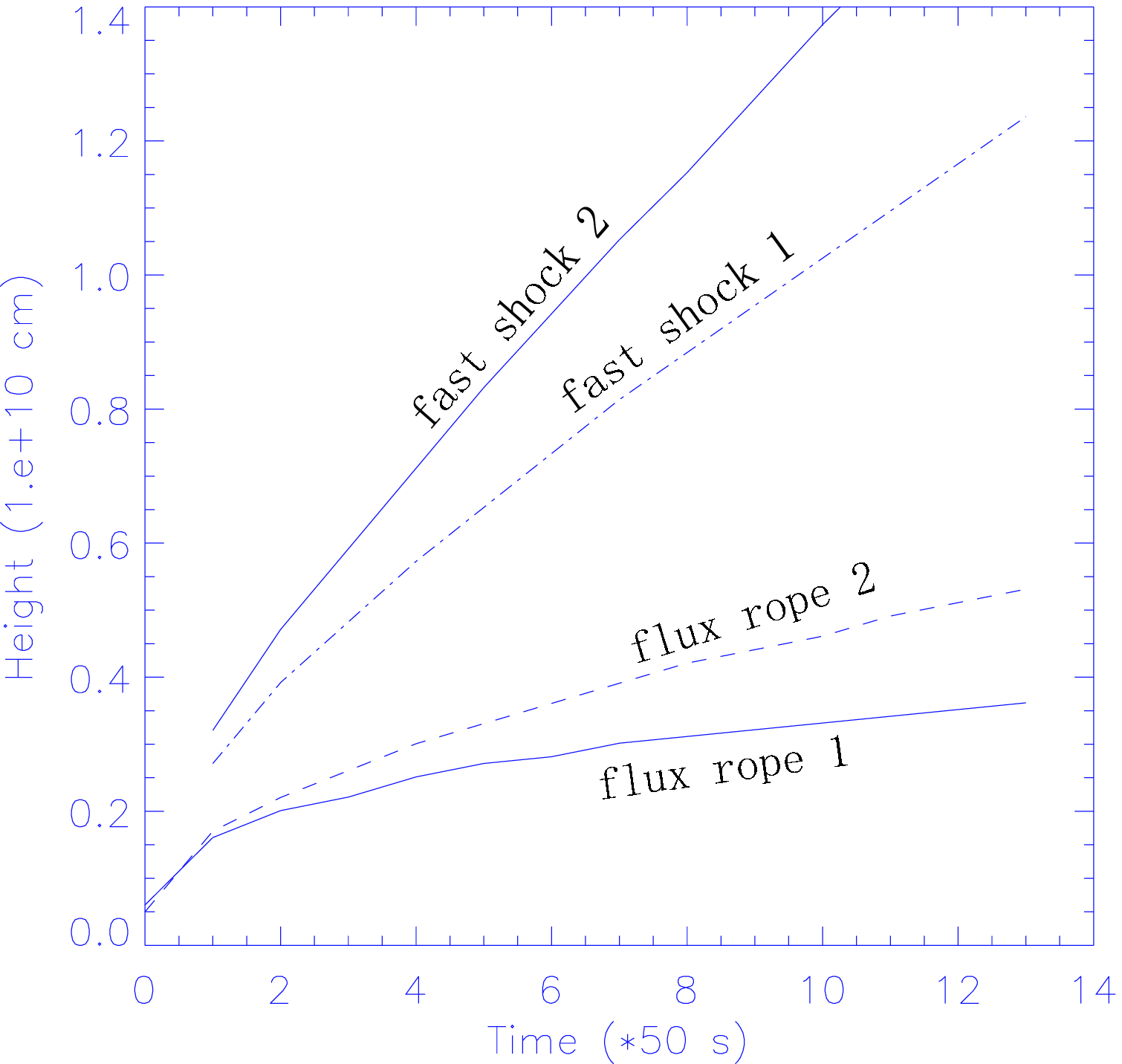}
\caption{Variations of the flux rope height and the height of the fast shock front versus time for two cases: Curves flux rope 1 and fast shock 1 are for the case without including the gravity; and curves flux rope 2 and fast shock 2 for the case including the gravity.}
\label{fig:h}
\end{figure}

Figure \ref{fig:h} displays the heights of the flux rope and the
fast mode shock versus time for two cases. Curves flux rope 1 and
fast shock 1 are for the case without including the gravity; and
curves flux rope 2 and fast shock 2 for the case including the
gravity. Obviously, curves in two cases have nearly the same form,
but those in case 2 manifest apparently more energetic behavior.
This confirms the conclusion of Lin \& Forbes (2000) and Lin (2002)
that the flux rope in the isothermal environment is much more easily
to escape following the catastrophe than in the constant density
environment.

\begin{figure}
\centering
\includegraphics[width=14cm,clip,angle=0]{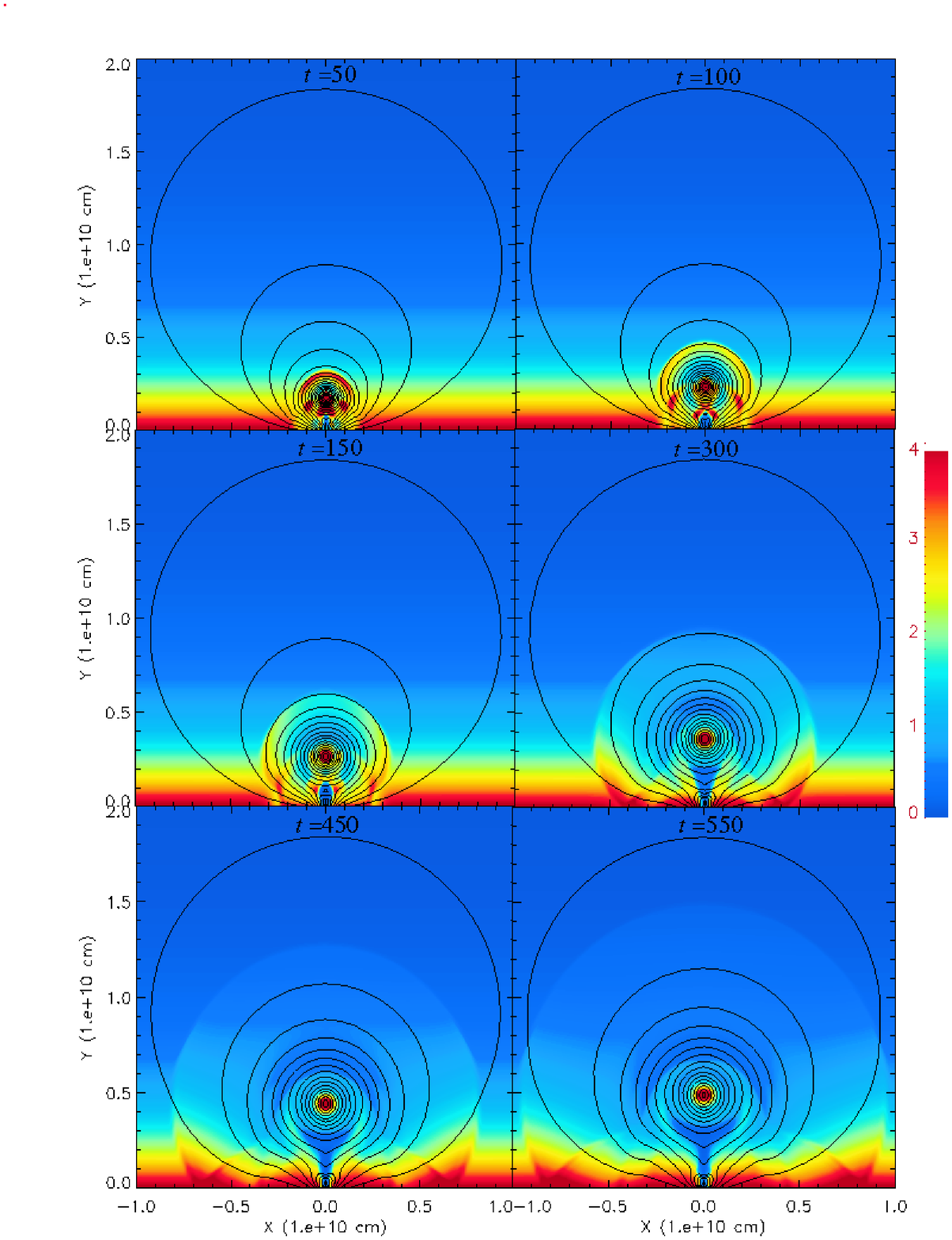}
\caption{Evolutions of the magnetic field and the plasma density as the flux rope moves outward in the case including the gravity. Continuous contours are magnetic field lines and the color shadings show the density distribution. The unit of time is second. The right color bar represents values of the density.}
\label{fig:gart}
\end{figure}

\begin{figure}
\centering
\includegraphics[width=16cm,clip,angle=0]{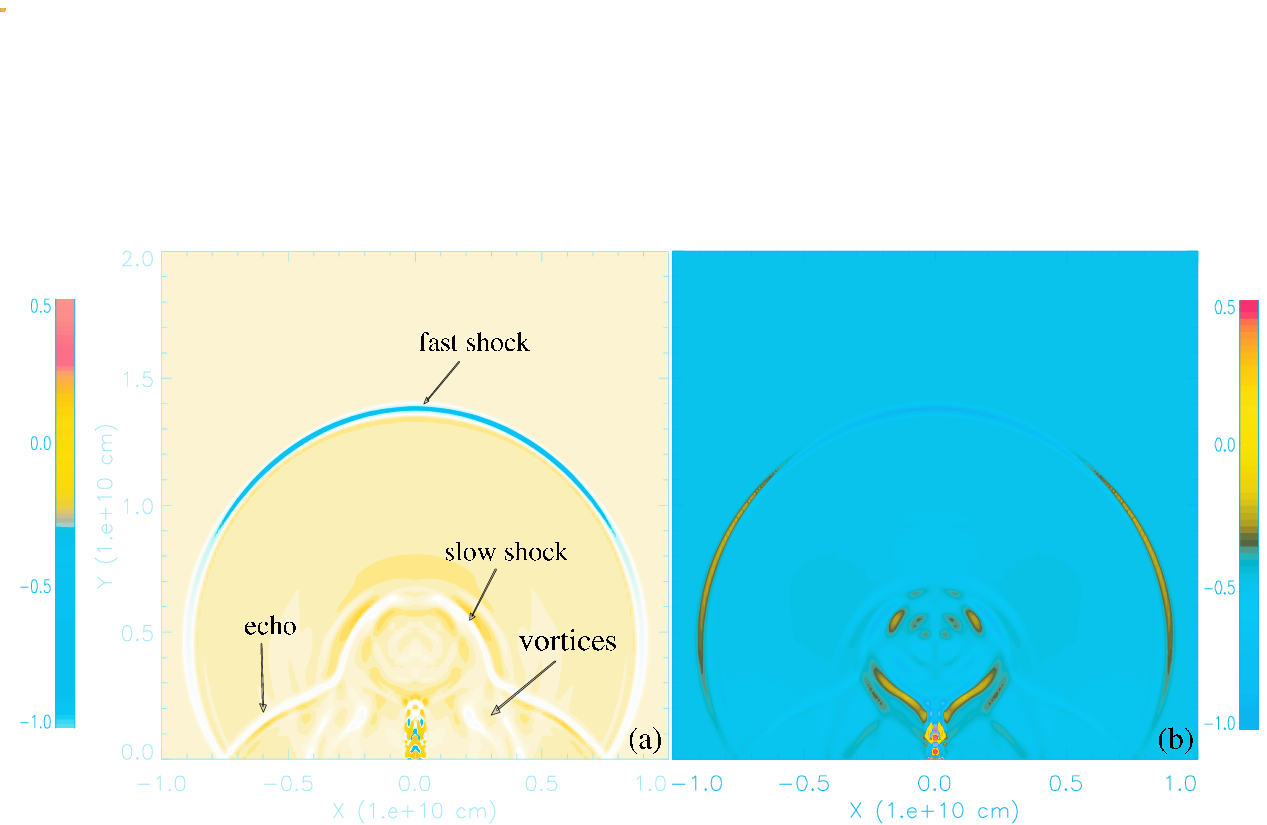}
\caption{Contours of $\nabla\cdot {\bf v}$ (a) and $(\nabla\times
{\bf v})_{z}$ (b) at $t=500$~s as the eruption progresses in the
environment including the gravity. The color shadings represent the
distribution of values of $\nabla\cdot {\bf v}$ and $(\nabla\times
{\bf v})_{z}$. The fast shock, slow shock, echo and vortices are
denoted in (a). In panel (b), a pair of the Petschek slow shocks
developing from the X-point is very impressive. The left and the
right color bars represent, in arbitrary unit, the values of
$\nabla\cdot {\bf v}$ and $(\nabla\times {\bf v})_{z}$,
respectively.} \label{fig:gd500}
\end{figure}

In addition, the motion of the flux rope also yields various disturbance phenomena nearby as expected. Figure \ref{fig:gart} plots the magnetic field and the plasma density at different stages in this process. We see from the snapshots in Figure \ref{fig:gart} the fast mode shock forming in front of the flux rope, expanding sideward and backward , and producing a crescent structure around the flux rope. At about $t=150$~s, it touches the boundary surface, and produces echoes propagating backward into the corona. The light yellow region nearby the bottom boundary $y=0$ corresponds to the footprints of the echo, which can be clearly recognized in the time interval between $t=150$~s and $t=550$~s. Comparing with Figure \ref{fig:ad}, we notice that the main evolutionary features in the system are duplicated, but an important difference is seen clearly: more significant disturbance with more complex patterns in the region near the bottom boundary are produced. Here to show detailed features and properties of the disturbance in this region, we adjust the color code for density distribution such that the fast shock is not well displayed in the panels of late time ($t> 300$ s). But keep in mind that the fast shock is there as shown by Figure \ref{fig:gd500}. We see that the density distribution near the bottom manifest apparent wave-like behavior, and such behavior varies with time. This reminds us the oscillatory behaviors of the EIT wave observed by Ballai et al. (2005) and Long et al. (2008).

We also investigate the divergence and the vorticity of the velocity, $\nabla\cdot {\bf v}$ and $(\nabla\times {\bf v})_{z}$, respectively, and their evolution in the eruptive process. Figure \ref{fig:gd500} shows the magnitude shadings of $\nabla\cdot {\bf v}$ (panel a) and $(\nabla\times {\bf v})_{z}$ (panel b) at $t=500$~s. The color shadings represent the distribution of values of $\nabla\cdot {\bf v}$ and $(\nabla\times {\bf v})_{z}$ in space, and several important features are denoted in panel (a). The fast shock, the slow shock, and the velocity vortices can be seen clearly from these panels.

Figure \ref{fig:gyv} plots the distribution of the $z$-component of
$\nabla\times{\bf v}$ at various times. As we did for Figure
\ref{fig:yv}, panels (a) and (b) show the distributions of
$(\nabla\times{\bf v})_{z}$ on the boundary surface $y=0$, and at
layer $y=0.3$, respectively. We see the similar propagating features
on two layers, but the speed is higher than in previous case. The
information revealed by Figure \ref{fig:gyv}a suggests that the fast
shock sweeps the bottom boundary surface at speed of around
138~km~s$^{-1}$, comparing with 126~km~s$^{-1}$ in the case of
constant density. The sign of the slow shock or the velocity
vortices at layer $y=0.0$ is not clear as expected. In Figure
\ref{fig:gyv}b, on the other hand, there are two propagating
features present, the faster one results from the fast shock in the
corona, and the slower one should be the footprint of either the
slow shock or the velocity vortices or the both in the corona. The
speed of the slow propagation is about 67~km~s$^{-1}$, comparing
with 50~km~s$^{-1}$ in the constant density case. Therefore,
including the gravity does not change our conclusions obtained on
the basis of studies in previous sections about the origin and
property of the Moreton wave and the EIT wave, and the
stratification of the atmosphere due to the gravity yields more
wave-like patterns in the lower corona as shown by Figure
\ref{fig:gart} comparing with Figure \ref{fig:ad}.

\section{Discussions}
Following the work of Forbes (1990), we used the ZEUS-2D code to perform a set of numerical experiments for the dynamical properties of several phenomena associated with CMEs in the eruption. In our study, the magnetic configuration includes a current-carrying flue rope that models the filament, and the background field is equivalent to that produced by a line dipole located in the photosphere. The MHD equations applied to describe the process of interest are physically ideal, but in the numerical code that performs calculations on the discrete grids, the numerical diffusion was inevitably invoked. We did not bother to remove this artificial effect since there is not an existing approach for the time being that could eliminate the numerical diffusion properly, and this work mainly focuses on the disrupting magnetic field and the associated phenomena.

\begin{figure}
\centering
\includegraphics[width=16cm,clip,angle=0]{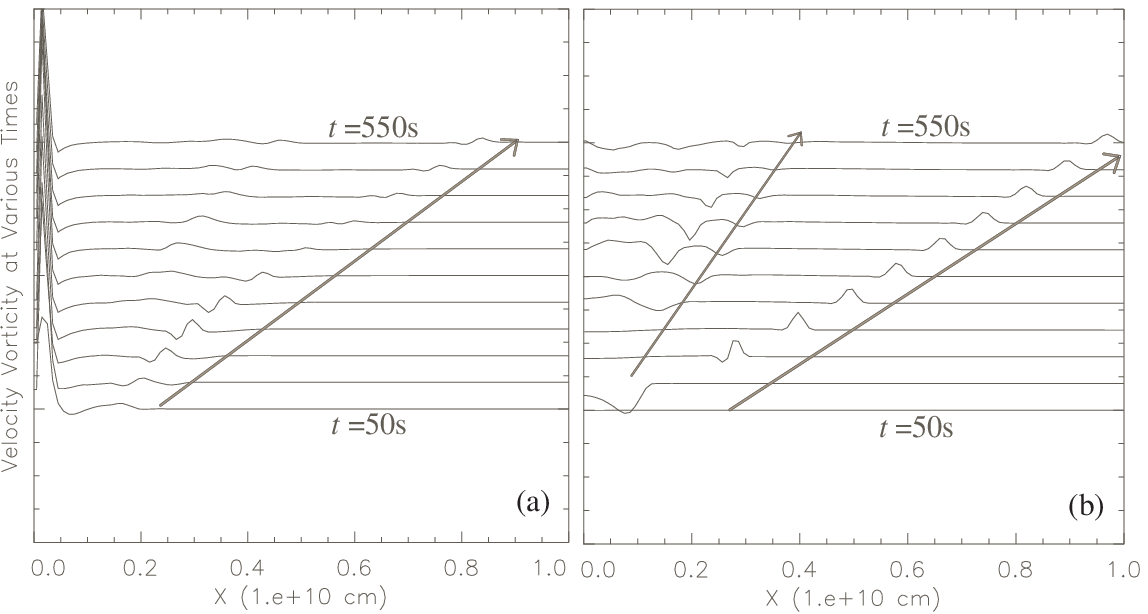}
\caption{Distributions of $(\nabla\times{\bf v})_{z}$ on layers $y=0.0$ (a), and  $y=0.3$ with the gravity included. The time intervals for drawing these curves in each panel are between 50~s and 550~s.}
\label{fig:gyv}
\end{figure}

We start the experiment with the state at which the system is not in equilibrium. So the flux rope rises rapidly at the very beginning. No current sheet forms from the reconnection region because of the numerical diffusion. Although such a diffusion is somewhat artificial, it does not influence our results and conclusions.

As the flux rope propagates, a fast mode and a slow mode shocks may be produced in front of it, a dimming area around the flux
rope and the reconnection region appears and can be easily
recognized. As impact of the fast mode shock reaches the boundary
surface, a pair of echoes developed from the footpoints of the fast
shocks. But the slow mode shock decays quickly and totally dissipated somewhere in the corona, and does not reach the bottom boundary. In this process, magnetic field lines are apparently deformed by the velocity vortices as a result of the the flux rope motion (see Figure \ref{fig:vortcs} and also Figure 10 of Forbes 1990).

More details of these features are revealed by the velocity
divergence $\nabla\cdot {\bf v}$. Contour plots of $\nabla\cdot {\bf
v}$ show more clearly the expansion of both the fast and the slow
shocks, as well as the echo formation with the fast shock reaching
the bottom boundary. With its continuous expansion, the fast shock
sweeps the boundary surface at its footprints, manifesting a
scenario of wave propagation at high speed, which reminds us the
Moreton wave observed in the chromosphere.

The plot also shows very complex features in the region where the
slow shocks mix with the velocity vortices behind the flux rope. As noted by Forbes (1990), this region propagates outward roughly in horizontal direction as the flux rope moves upward. But as shown by Figure \ref{fig:y0d}, no sign of either the vortices or the slow shock could be recognized on the bottom boundary. This implies that the vortices and the slow shock cannot cause disturbance to the lower atmosphere. Behaviors of the velocity vorticity $\nabla\times {\bf v}$ further confirms this point.

Plot of $(\nabla\times {\bf v})_{z}$ versus $x$ on the boundary surface of
$y=0$ displays the propagating feature of the fast shock only
(Figures \ref{fig:yv}a and \ref{fig:gyv}a), but we see the slow shock disturbance at layer of $3\times 10^{4}$~km above the boundary surface (Figures \ref{fig:yv}b and \ref{fig:gyv}b), which is roughly 0.3 scale height of the domain investigated. We also tried to explore signs of the slow shock and the velocity
vortices in a lower layer, say of $2\times 10^{4}$~km, but no sign
was found, which is suggestive of disappearance of the impact of the
slow shock and the velocity vortices in lower layers of the atmosphere. This result, together with the speed the slow shock ($\sim$ 40\% the fast shock speed) sweeps the corona horizontally, relates the EIT wave origin directly to the slow shock since observations yield almost the same result for altitude and the speed at which the EIT wave appears and propagates. Wills-Davey et al. (2007) went through existing observational results and investigated new events, and they found that the layers where the EIT wave is most likely to appear are of the altitude of roughly 0.3 scale height of the coronal environment; Thompson \& Myer (2009) showed that the EIT wave speed is usually $1/5\sim 1/3$ that of Moreton wave.

In addition to assisting identifying the origin of the Moreton wave
and that of the EIT wave, the above results is also helpful for
understanding several correlations usually observed in solar
eruptions. These correlations include that of the EIT wave to the
Moreton wave, that of the EIT wave to the type II radio burst, and
that of the Moreton wave to the type II burst. Our results obtained
here from the numerical experiment indicate the origin of the
Moreton wave from the fast mode shock driven by the flux rope (CME),
and both observations and theories suggest the CME-driven shock
origin of the type II radio burst as well (e.g., see Lin et al. 2006
for a brief review). Therefore, a good correlation of the Moreton
wave to the type II radio burst is expected as indicated by many
observations (e.g., see Wild et al. 1963; Kai 1970; Uchida 1974;
Pinter 1977; Cliver et al. 1999; Chen et al. 2005a). On the other
hand, the EIT wave has different original source; it thus does not
necessarily correlate to the other two phenomena in an obvious
fashion. This may account for the poor correlation of the EIT wave
to the type II radio burst revealed by observations (e.g., see
Klassen et al. 2000 and Chen et al. 2005a).

\section{Conclusions}
The numerical experiment provides us an approach to studying many
evolutionary details in the disrupting magnetic field that could not
be investigated in the framework of the analytic solution. As
indicated by the analytical model of CMEs (e.g., see Lin \& Forbes
2000 and references therein), the flux rope that is used to model the filament is thrust outward rapidly as the magnetic configuration loses the mechanical
equilibrium in a catastrophic fashion. Fast motion of the flux rope
quickly produces two types of shocks in front of it, and a pair of
velocity vortices behind it. We find for two cases that consequences
of these shocks and vortices could well correspond to several
important observational phenomena, including the Moreton wave, EIT
waves, coronal dimmings, and so on. Some issues that are still of
active research topics are also discussed. The main results are
summarized as follows.

1. As expected, a dimming region around the flux rope in the
disrupting magnetic field forms. Its area expands in the eruptive
process. Studying the plasma properties indicates that the density
in this area is apparently lower than the surroundings, and further
confirms that the dimming results from the depletion of the material
in the relevant region.

2. A fast mode shock forms forward of the flux rope as the eruption
progresses. The shock expands toward its flank when propagating
forward, and eventually reaches the lower layer of the atmosphere,
such as the chromosphere. An echo develops from the footpoints as a
result of the shock being reflected from the boundary surface.
Interaction between the fast shock and the lower layer results in
disturbance, which accounts for the Moreton wave observed in
H$\alpha$.

3. A slow model shock is created as well at both sides of the flux rope when the fast shock forms, but leaves farther and farther behind the fast shock in the propagation. It also expands toward its flank, but its impact cannot reach the bottom boundary. Instead the disturbance it causes vanishes somewhere above the boundary.

4. We also investigated the velocity vortices, and found that it
usually forms slightly below the reconnection region, where the slow
shock is able to reach. Our results suggest that the joint impact of
the slow shock and the vortices is quite likely to account for the
EIT waves.

5. The CME-driven fast shock is the sole source of the metric type
II radio burst, and the Moreton wave is also a consequence of the
CME-driven shock, not the flare-initiated blast wave. The same
igniter of the Moreton wave and the type II burst explains the good
correlation of these two phenomena to one another (Cliver et al.
1999). This, together with the fact that the EIT wave is either the
slow shock, or the velocity vortices, or both, in origin, further
accounts for the lack of correlation of type II radio bursts to EIT
waves (Klassen et al. 2000; Chen et al. 2005a).

6. We also investigated the case in which the gravity is included, and found that existing of the gravity does not change our main results deduced from the case without the gravity. But the stratification of the atmosphere due to the gravity yields much more complicated patterns of the disturbance created by shocks and the velocity vortices. Such disturbance in the layer above the bottom boundary displays apparent wave-like features. Considering that the EIT wave appears roughly in the same layer, we conclude that the wave-like feature revealed by our results is quite likely to account for the oscillatory property of the EIT wave observed by Ballai et al. (2005) and Long et al. (2008).

\acknowledgments Authors appreciate M. Zhang very much for teaching
the key techniques of running the ZEUS-2D code smoothly. They are
also grateful to T. G. Forbes, P. Chen, G. D. R. Atrill, M. J.
Wills-Davey, and the referee for valuable comments and suggestions for helping improve this paper. This work was supported by Program 973 grant 2006CB806303, by NSFC grants 40636031 and 10873030, and by CAS grant KJCX2-YW-T04.  JL's work at CfA was supported by NASA grant NNX07AL72G when he visited to CfA.

\end{document}